\documentclass[twocolumn]{aastex61}

\usepackage[T1]{fontenc}
\usepackage{ae,aecompl}

\usepackage{graphicx}	
\usepackage{amsmath}	
\usepackage{amssymb}	
\usepackage{bm}

\received{XXX}
\revised{XXX}
\accepted{XXX}

\shorttitle{Full-sky ray-tracing simulation}
\shortauthors{C.Wei et al.}

\begin{document}

\title{Full-sky ray-tracing simulation of weak lensing using ELUCID simulations: \\ exploring galaxy intrinsic alignment and cosmic shear correlations}

\author{Chengliang Wei}
\affiliation{Purple Mountain Observatory, the Partner Group of MPI f\"ur Astronomie,  2 West beijing Road, Nanjing 210008, China; \href{mailto:Guoliang@pmo.ac.cn}{Guoliang@pmo.ac.cn}; \href{mailto:Kangxi@pmo.ac.cn}{Kangxi@pmo.ac.cn}}
\affiliation{Graduate School, University of the Chinese Academy of Science, 19A, Yuquan Road, Beijing 100049, China}

\author{Guoliang Li}
\affiliation{Purple Mountain Observatory, the Partner Group of MPI f\"ur Astronomie,  2 West beijing Road, Nanjing 210008, China; \href{mailto:Guoliang@pmo.ac.cn}{Guoliang@pmo.ac.cn}; \href{mailto:Kangxi@pmo.ac.cn}{Kangxi@pmo.ac.cn}}

\author{Xi Kang}
\affiliation{Purple Mountain Observatory, the Partner Group of MPI f\"ur Astronomie,  2 West beijing Road, Nanjing 210008, China; \href{mailto:Guoliang@pmo.ac.cn}{Guoliang@pmo.ac.cn}; \href{mailto:Kangxi@pmo.ac.cn}{Kangxi@pmo.ac.cn}}

\author{Yu Luo}
\affiliation{Purple Mountain Observatory, the Partner Group of MPI f\"ur Astronomie,  2 West beijing Road, Nanjing 210008, China; \href{mailto:Guoliang@pmo.ac.cn}{Guoliang@pmo.ac.cn}; \href{mailto:Kangxi@pmo.ac.cn}{Kangxi@pmo.ac.cn}}

\author{Qianli Xia}
\affiliation{Purple Mountain Observatory, the Partner Group of MPI f\"ur Astronomie,  2 West beijing Road, Nanjing 210008, China; \href{mailto:Guoliang@pmo.ac.cn}{Guoliang@pmo.ac.cn}; \href{mailto:Kangxi@pmo.ac.cn}{Kangxi@pmo.ac.cn}}
\affiliation{Graduate School, University of the Chinese Academy of Science, 19A, Yuquan Road, Beijing 100049, China}

\author{Peng Wang}
\affiliation{Purple Mountain Observatory, the Partner Group of MPI f\"ur Astronomie,  2 West beijing Road, Nanjing 210008, China; \href{mailto:Guoliang@pmo.ac.cn}{Guoliang@pmo.ac.cn}; \href{mailto:Kangxi@pmo.ac.cn}{Kangxi@pmo.ac.cn}}
\affiliation{Graduate School, University of the Chinese Academy of Science, 19A, Yuquan Road, Beijing 100049, China}

\author{Xiaohu Yang}
\affiliation{Department of Astronomy, Shanghai Jiao Tong University, Shanghai 200240, China}

\author{Huiyuan Wang}
\affiliation{Key Laboratory for Research in Galaxies and Cosmology, Department of Astronomy, University of Science and Technology of China, Hefei, Anhui 230026, China}

\author{Yipeng Jing}
\affiliation{Department of Astronomy, Shanghai Jiao Tong University, Shanghai 200240, China}

\author{Houjun Mo}
\affiliation{Department of Astronomy, University of Massachusetts, Amherst MA 01003-9305, USA}

\author{Weipeng Lin}
\affiliation{School of Physics and Astronomy, Sun Yat-Sen University, Guangzhou 510275, China}

\author{Yang Wang}
\affiliation{School of Physics and Astronomy, Sun Yat-Sen University, Guangzhou 510275, China}

\author{Shijie Li}
\affiliation{Shanghai Astronomical Observatory, Nandan Road 80, Shanghai 200030, China}

\author{Yi Lu}
\affiliation{Shanghai Astronomical Observatory, Nandan Road 80, Shanghai 200030, China}

\author{Youcai Zhang}
\affiliation{Shanghai Astronomical Observatory, Nandan Road 80, Shanghai 200030, China}

\author{S.H. Lim}
\affiliation{Department of Astronomy, University of Massachusetts, Amherst MA 01003-9305, USA}

\author{Dylan Tweed}
\affiliation{Department of Astronomy, Shanghai Jiao Tong University, Shanghai 200240, China}

\author{Weiguang Cui}
\affiliation{Departamento de F\'isica Te\'orica, M\'odulo 15, Facultad de Ciencias, Universidad Aut\'onoma de Madrid, E-28049 Madrid, Spain}



\begin{abstract}

The   intrinsic  alignment of  galaxies  is  an  important  systematic
effect  in  weak-lensing surveys,  which  can  affect the derived
cosmological parameters.   One  direct way  to  distinguish  different
alignment models  and quantify their  effects on the measurement  is to
produce mocked  weak-lensing surveys.  In  this work, we  use full-sky
ray-tracing  technique to  produce mock  images of  galaxies  from the
ELUCID $N$-body simulation run with  the WMAP9 cosmology. In our model
we assume that the shape  of central elliptical galaxy follows that of
the dark matter  halo, and spiral galaxy follows  the halo spin. Using
the mocked galaxy images, a  combination of galaxy intrinsic shape and
the gravitational  shear, we  compare the predicted  tomographic shear
correlations to the results of  KiDS and DLS.  It is  found that our
predictions stay between the KiDS  and DLS results.   We rule  out a
model in  which the satellite  galaxies are radially aligned  with the
center galaxy,  otherwise the  shear-correlations on small  scales are
too  high.   Most  important,  we  find that  although  the  intrinsic
alignment  of  spiral   galaxies  is   very   weak,   they  induce  a
positive correlation  between the  gravitational shear signal  and the
intrinsic galaxy orientation (GI).   This is because the spiral galaxy
is  tangentially  aligned  with  the nearby  large-scale  overdensity,
contrary  to the radial  alignment of  elliptical galaxy.  Our results
explain  the origin of  detected positive  GI term  from the  weak-lensing
surveys. We conclude  that in  future  analysis,  the GI  model  must
include  the dependence on galaxy types in more detail.
\end{abstract}

\keywords{Gravitational lensing: weak --- Cosmology: large-scale structure in the universe --- Methods: numerical}




\section{Introduction}
In  the context of General  Relativity, photos  emitted from
distant  galaxies are continuously  deflected by  the intervening
mass field  of the large-scale  structures \citep{1992grle.book.....S,
  2006glsw.conf.....M,  2016arXiv161206535B}. This
gravitational lensing effect, referred to as ``cosmic shear", produces
some coherent  distortions of the  observed galaxy images, 
which can be measured to  probe  the matter  distribution in  the universe
\citep{1999ARA&A..37..127M,  2001A&A...374..757V, 2003astro.ph..9482K,
  2008A&A...479....9F, 2015RPPh...78h6901K, 2016MNRAS.463.3326F}.  
Great progress  has been made  in using   cosmic  shears  to  constrain   
cosmological  models (see \citealt[][]{2015RPPh...78h6901K} for a review), to estimate the  dark
energy parameter  $w$ \citep{2007NJPh....9..444B, 2009JCAP...06..026L,
  2015JCAP...04..048B}, and to  test theories  of modified
gravity \citep{2015PhRvD..92f4024L, 2016MNRAS.459.2762H}.

 Observational  results  from recent  weak-lensing
surveys,     such      as     Canada-France-Hawaii Telescope Lensing Survey
\citep[CFHTLenS, ][]{2012MNRAS.427..146H,  2013MNRAS.432.2433H}
        and        Deep       Lens        Survey
\citep[DLS, ][]{2013ApJ...765...74J, 2016ApJ...824...77J, 2016AAS...22730707J},
demonstrate that cosmic shears can be combined with other observations, 
such as the cosmic microwave background (CMB), baryon acoustic
oscillations (BAO), and galaxy cluster abundance, to break the degeneracy  
among different cosmological  parameters (e.g., $\Omega_{\rm m}$-$\sigma_{8}$).  
Thus,  accurate measurement of weak lensing effects has been  one of the main 
goals of  many ongoing  and upcoming galaxy  surveys,     such     as     the     Kilo-Degree     Survey
\citep[KiDS,][]{2015A&A...582A..62D},      Dark     Energy      Survey
\citep[DES,][]{2016MNRAS.460.1270D},    Hyper    Suprime-Cam    Survey
\citep[HSC,              ][]{2012SPIE.8446E..0ZM},              Euclid
\citep{2011arXiv1110.3193L}  and the  Large Synoptic  Survey Telescope
\citep[LSST, ][]{2009arXiv0912.0201L}.

These surveys will provide high-quality data  with very wide
sky coverages, and the statistical  uncertainties in weak lensing 
measurements are expected to be small. However,
high accuracy analyses of cosmic shear also requires 
understanding the systematics in  the data, such as  those in
measurements of galaxy ellipticity and the point spread function. 
In addition, accurate theoretical modeling is also necessary in order to 
interpret the  observed  data.  

One  of  the  most  serious  astrophysical  systematic
effects in  the era of  accurate weak-lensing analyses is the
intrinsic  alignment   (IA)  of  galaxies  (\citealt{2000MNRAS.319..649H,
  2001ApJ...559..552C,   2002MNRAS.335L..89J}; see
  \citealt{2015SSRv..193..139K,  2015PhR...558....1T} for a review), 
 which can  mimic the  gravitational  lensing  signal, thereby contaminating  
 the measurements     of      cosmic     shears.     
A significant signal of intrinsic alignments has been detected by 
\citet{2006MNRAS.367..611M} in the luminous  red galaxies from 
the Sloan Digital Sky Survey \citep[SDSS,][]{2000AJ....120.1579Y}, and 
the authors concluded that neglecting such alignments can lead to an underestimate 
of the linear amplitude of density fluctuations by $20\%$ for cosmic  
shear surveys at $z \sim 1$. Clearly, an accurate model  for galaxy IA, 
which is capable of describing its dependence on redshift and  galaxy 
properties,  is crucial for maximizing the science returns of 
ongoing and future weak lensing surveys \citep{2016MNRAS.456..207K}.

There have  been numerous investigations on  galaxy IA.
Based   on  the   tidal   field  theory   \citep{2001MNRAS.320L...7C},
\citet{2004PhRvD..70f3526H} developed  a linear  model for  galaxy IA,
which     was     later      improved   to include  some non-linear effects
\citep{2007NJPh....9..444B,  2012JCAP...05..041B}.  A  useful formula
with   a   single  parameter   was   introduced   by \citet{2011A&A...527A..26J},
which  can easily be included  in  the analyses     of    cosmic     shears    
from   observational  data
\citep{2010MNRAS.408.1502K,  2013MNRAS.432.2433H, 2016ApJ...824...77J,
  2017MNRAS.465.1454H,  2017MNRAS.465.2033J}.    
As a more accurate description of  galaxy IA, a halo model 
is developed \citep{2010MNRAS.402.2127S}, which can predict the
IA signal as a function of  galaxy properties.  However, as pointed in
\citet{2013MNRAS.431..477J},   most  of these simple  IA  models   
are  expected to work only at low $z$, and it is  still unclear  how 
galaxy  IA varies  as a function of galaxy properties at high $z$.

$N$-body and  Hydro-dynamical simulations  are also extensively  used to
study galaxy  IA.  When $N$-body  simulations are used for the purpose, 
assumptions about the connection  between galaxy shape and dark  matter halo shape
have  to be  made.  \citet{2007MNRAS.378.1531K} used  $N$-body  simulations to
explain  the  observed small-scale  alignment  of satellites  galaxies
around  central galaxies  in the  SDSS data  \citep{2006MNRAS.369.1293Y}. 
They found that the orientations of elliptical galaxies follow that of the host halos, 
albeit with some mis-alignment, and that the spins of spiral galaxies
follow that of their host halos.  This assumption is later confirmed 
\citep[e.g.,][]{2009RAA.....9...41F,
2009ApJ...694..214O, 2010ApJ...709.1321A}.   With similar assumptions
about how     galaxies are aligned with  dark    matter     halos,
\citet{2013MNRAS.436..819J}  measured galaxy IA  on large  scales from
the Millennium Simulations  \citep{2005Natur.435..629S} and found that
early-type  galaxies are  strongly aligned  with each  other,  but
spiral galaxies do not show significant correlation signals between their intrinsic 
ellipticities. This dependence on galaxy-type agrees with observational results.
\citep[e.g.,][]{  2011A&A...527A..26J,
  2013MNRAS.432.2433H,  2011MNRAS.410..844M}.   More recently, cosmological
hydrodynamical simulations have been used to predict the galaxy
IA    \citep[e.g.,][]{2014ApJ...791L..33D,      2014MNRAS.441..470T,
  2015MNRAS.454.2736C,    2016MNRAS.461.2702C,    2017MNRAS.472.1163C,
  2015MNRAS.454.3328V, 2017MNRAS.468..790H, 2016MNRAS.462.2668T}.  
The main merit of using a hydro-dynamical simulation is that galaxy shapes 
are directly  predicted by the simulation.  In agreement  with previous
analytical models  and $N$-body simulations,  these hydro-dynamical
simulations also indicate that elliptical  galaxies have a stronger tendency
to align with each other on large scales than do spiral galaxies.
However,  due to  limited  volumes of  these  simulations (often  around
100Mpc/$h$) and different treatment of baryonic physics, the predicted
galaxy IA  signal and its dependence on galaxy  properties and redshift  
still varies from simulation to simulation.

Although an  accurate   model  for  galaxy   IA  is  still  not   available  at
the present, the  main   assumption,   adopted  in   $N$-body
simulations,  that elliptical  galaxies follow the shapes, 
while spirals    follow   the   spins,   of   host halos
\citep[e.g.,][]{2013MNRAS.436..819J},  can  be  checked by
comparing real and  mocked observational data of galaxy shear
correlations. This can be achieved  by using ray tracing in an $N$-body
simulation combined with a model of galaxy formation which can predict
galaxy shapes, luminosities and positions.  With such an approach, we can
produce  observable images  of   galaxies  and  obtain  the  auto-  and
cross-correlation  functions between  gravitational  shear and  galaxy
intrinsic ellipticity at different redshifts.  We can then compare 
model predictions with results obtained from two recent surveys, KiDS and 
DLS, and examine the importance of galaxy IA.  The results of these two
surveys show a $\sim 2\sigma$  tension in 
$S_{8}  \equiv \sigma_{8} \sqrt{\Omega_{\rm  m}/0.3}$,  
with  KiDS giving $S_{8}  =  0.745  \pm  0.039$  and DLS giving 
$0.818_{-0.026}^{+0.034}$. The main goal of this paper is to use 
such approach to constrain galaxy IA models and to examine 
the contamination from IA in the 2-point correlation functions 
of the cosmic shear.

As a `standard' algorithm, the multiple-plane ray-tracing simulation
with the flat-sky approximation  
\citep[e.g.,][]{2000ApJ...530..547J,  2004APh....22...19W, 2009A&A...499...31H}
has been widely used to simulate lensing
maps for small-field survey. It also roughly works for hundreds of square degree 
surveys, such as KiDS with 450 square degrees \citep[hereafter KiDS-450,][]{2017MNRAS.465.1454H}, 
but will not suitable for even large-filed surveys such as Euclid and LSST \citep{2016arXiv161104954K, 2017arXiv170205301K, 2017JCAP...05..014L}. 
To quantify the effect of cosmic variance in the small-field surveys, 
one needs construct a lot of light cones to simulated different realizations.  
In this paper, we adopt ray-tracing code on a curved sky to simplify 
this procedure and to prepare for these large-field surveys.

Full-sky weak-lensing maps have already been constructed in
a number of papers   
\citep{2009A&A...497..335T,    2013MNRAS.435..115B,
  2008MNRAS.391..435F,                            2015MNRAS.447.1319F,
  2015MNRAS.453.3043S}. These  simulations usually cover a  sufficiently large
volume to compute a full-sky convergence (and shear) maps, and explore
the lensing power at  both the linear and  nonlinear regimes.  
In this paper, we follow the  ray-tracing  method of
\citet{2008ApJ...682....1D, 2009A&A...497..335T, 2013MNRAS.435..115B}.
We perform high resolution (both in space and  in mass) lensing simulations,
using an  iterative scheme  of spherical  harmonic analysis, to model   
lensed properties  of 'semi-analytic' galaxies  in the simulation.
These simulated galaxies allow us to study the statistical properties
of  galaxy alignments,  and to compare  our mock  observations with  the
observational  results  from both DLS  \citep{
  2016ApJ...824...77J,      2016AAS...22730707J}     and      KiDS-450
\citep{2017MNRAS.465.1454H} using tomographic correlation functions.

The paper  is organized  as follows. In  Section 2 we first  summarize the
basic theoretical background of weak  lensing, focusing on the power
spectrum and shear correlation analyses.  In Section 3 we introduce 
the simulations and the  spherical ray-tracing technique. Section 4 describes
how we model galaxy properties, such as  luminosity, morphology and shape, from the
semi-analytic  model, and we also present  results of 
intrinsic  alignments  of galaxies and their dependence on  galaxy type 
and halo  mass.  In Section 5,  we describe the tomographic  analyses of cosmic  
shears in  our lensing  simulation, compare model predictions with 
observational data,  and quantify the contributions of the 
intrinsic-intrinsic (II) shear correlation and the gravitational  shear-intrinsic  
(GI) shear  correlation by spiral  and elliptical  galaxies.  Conclusions and  
discussions  are given in Section 6.


\section{COSMOLOGICAL WEAK LENSING}

In  this section,  we briefly summarize the  theoretical  background for the
analyses  of weak gravitational  lensing and  describe some  basics about
intrinsic alignment and shear correlations.

\subsection{Basics}

In general,  for a  source galaxy with  the observed  angular position
$\bm{\theta}$  and  its real  position  $\bm{\beta}$,  one  can
characterize the  deformation effect of  cosmic shear through
the    distortion   matrix    \citep{1992grle.book.....S,
  2000ApJ...530..547J},
\begin{equation} \label{equ:matrixA}
  \mathcal{A}(\bm{\theta})=\frac{\partial\bm{\beta}}{\partial\bm{\theta}}
  \equiv\begin{pmatrix}1-\kappa-\gamma_1 & -\gamma_2-\omega \\ -\gamma_2+\omega & 1-\kappa+\gamma_1 \end{pmatrix},
\end{equation}
where          $\kappa$          is          the          convergence,
$\bm{\gamma}=\gamma_1+{\rm i}\gamma_2$  defines the complex  shear in
lensing, and the  additional antisymmetric quantity, $\omega$, describes
an overall rotation in the  lensed images.  In the weak lensing regime
(i.e.,  $\kappa,\bm{\gamma} \ll  1$) and to the  linear  order, 
the components  of the matrix  are related to  the  second  derivatives  of  the
gravitational       potential      as      \citep{2001PhR...340..291B,
  2009A&A...499...31H, 2015RPPh...78h6901K}
\begin{equation}
  \mathcal{A}_{ij}(\bm{\theta}, \chi)=\delta_{ij}-\frac{2}{c^2}\int_{0}^{\chi} {\rm d}\chi'\frac{r(\chi-\chi')r(\chi')}{r(\chi)}\Phi_{,ij}(r(\chi')\bm{\theta},\chi'),
\end{equation}
where $\delta_{ij}$  is the Kronecker delta,  $c$ is the  speed of 
light, $\chi$  is the comoving  distance and $r(\chi)$  the
comoving  angular   diameter  distance.   According   to  the  Poisson
equation, gravitational potential $\Phi$ can be related to the density
contrast $\delta$.   Hence, convergence $\kappa$  can be
expressed as a weighted integral  of the over-density $\delta$ along the line
of sight,
\begin{equation}
  \kappa(\bm{\theta}, \chi)=\frac{3H_{0}^{2}\Omega_{{\rm m}}}{2c^2}\int_{0}^{\chi}{\rm d}\chi'\frac{r(\chi-\chi')r(\chi')}{r(\chi)}\frac{\delta(r(\chi')\bm{\theta},\chi')}{a(\chi')},
  \label{equ:kappa}
\end{equation}
where $H_{0}$  is the Hubble  constant, $\Omega_{{\rm m}}$ is
the matter density in units of the critical density,
and $a(\chi')$ is the scale factor at $\chi'$.

\subsection{Power Spectrum of Weak Lensing Field}\label{sec:introPS}

In  the  flat-sky  limit,   the  power  spectrum  of  the  convergence
$C^{\kappa\kappa}(\ell)$  on the modulus $\ell$ is known  as the  two-point  
correlation in Fourier space,
\begin{equation}
  \langle\tilde{\kappa}(\bm{\ell}) \tilde{\kappa}^{*}(\bm{\ell}')\rangle=(2\pi)^2\delta_{\rm D}(\bm{\ell}-\bm{\ell}')C^{\kappa\kappa}(\ell),
\end{equation}
where  $\delta_{\rm D}(\bm{\ell})$ is  the Dirac  delta function.
Using  equation~(\ref{equ:kappa}),  one  can  derive the  angular  power
spectrum of the convergence field in the Limber approximation,
\begin{equation}
  C^{\kappa\kappa}(\ell) = \int^{\chi_{\rm H}}_{0} {\rm d}\chi \frac{W(\chi)^{2}}{r(\chi)^{2}} P_{\delta}\left(k=\frac{\ell}{r(\chi)}, \chi\right),
\label{eq:Ckk}
\end{equation}
where  $P_{\delta}(k,\chi)$ is the  3-D power  spectrum of  the matter
distribution at  the given comoving distance $\chi$,  and the integral
is  calculated  along  the  line  of  sight  to  the  comoving horizon distance
$\chi_{\rm H}$.  Here the weight function $W(\chi)$ is defined as,
\begin{equation}
  W(\chi) = \frac{3H_{0}^{2}\Omega_{{\rm m}}}{2c^2}\frac{r(\chi_{\rm H}-\chi)r(\chi)}{r(\chi_{\rm H})} \frac{1}{a(\chi)}.
\end{equation}
From the  non-linear theoretical  models, such as  the Halofit  model
\citep{2003MNRAS.341.1311S, 2012ApJ...761..152T},  one can predict the
convergence   power   spectrum   $C^{\kappa\kappa}(\ell)$  from   the
non-linear $P_{\delta}(k)$.  Therefore, the weak lensing survey can be
used to probe the gravitational growth of the density structure.

While dealing with full-sky measurements, it is useful to note that
the  angular   power  spectrum   of  weak  lensing convergence $\kappa$
and complex shear $\bm{\gamma}$  can be derived
from    the   spin-$s$    spherical    harmonics   $_{s}Y_{\ell}^{m}$
\citep{1996astro.ph..9149S}.  The relations of power spectra
between     the      convergence,     shear     E-      and     B-mode
\citep{2002A&A...389..729S,  2003NewAR..47..987B, 2003PhRvD..67b3501B,
  2010PhRvD..82b3001Z}        have        been       derived        by
\citet{2000PhRvD..62d3007H} for an all-sky lensing deformation tensor
field.   Here we  briefly  summarize the  spin-$s$ spherical  harmonic
decomposition  of  the  full-sky  lensing,  referring   the  reader  to
\citet{2000PhRvD..62d3007H}  for  detailed discussions  of  the  power
spectrum in weak lensing.

As reviewed in the appendix of \citet{2013MNRAS.435..115B}, the convergence,
lensing shear and rotation in the distortion matrix (Eq.~\ref{equ:matrixA}) can be
decomposed by the spherical harmonics \citep{2000PhRvD..62d3007H, 2013MNRAS.435..115B},
\begin{equation}
  \kappa(\hat{\bm{n}}) = -\frac{1}{2}\sum_{\ell m}\ell(\ell+1)\phi_{\ell m}Y_{\ell}^{m}(\hat{\bm{n}})
\end{equation}
\begin{equation}
  \gamma_{1}(\hat{\bm{n}}) \pm i\gamma_{2}(\hat{\bm{n}}) =
  \frac{1}{2}\sum_{\ell m}\sqrt{\frac{(\ell+2)!}{(\ell-2)!}} (\phi_{\ell m} \pm i \Omega_{\ell m}) _{\pm2}Y_{\ell}^{m}(\hat{\bm{n}})
\end{equation}
\begin{equation}
  \omega(\hat{\bm{n}}) = -\frac{1}{2}\sum_{\ell m}\ell(\ell+1)\Omega_{\ell m}Y_{\ell}^{m}(\hat{\bm{n}}),
\end{equation}
where $\phi$ is the lensing deflection potential, $\Omega$ is
the pseudo-scalar potential \citep[as described by][]{1996astro.ph..9149S},
and $\hat{\bm{n}}$ denotes a given position on the sky.
Consequently, the different power spectra can be related as,
\begin{equation}
  C^{\kappa\kappa}(\ell) = \frac{1}{4}\ell^2 (\ell+1)^2 C^{\phi\phi}(\ell)
\end{equation}
\begin{equation}
  C^{\omega\omega}(\ell) = \frac{1}{4}\ell^2 (\ell+1)^2 C^{\Omega\Omega}(\ell)
\end{equation}
\begin{equation}
  C^{EE}(\ell) = \frac{1}{\ell^2 (\ell+1)^2} \frac{(\ell+2)!}{(\ell-2)!} C^{\kappa\kappa}(\ell)
\end{equation}
\begin{equation}
  C^{BB}(\ell) = \frac{1}{\ell^2 (\ell+1)^2} \frac{(\ell+2)!}{(\ell-2)!} C^{\omega\omega}(\ell).
\end{equation}
Thus in the flat-sky limit, one has $\frac{1}{\ell^2 (\ell+1)^2} \frac{(\ell+2)!}{(\ell-2)!} \approx 1$, 
showing that $C^{EE}(\ell)$ equals $C^{\kappa\kappa}(\ell)$ at small scales \citep{2016arXiv161104954K, 2017arXiv170205301K}.

\subsection{Cosmic Shear and Intrinsic Alignment}

Weak  lensing will induce  an additional  coherent deformation  to the
intrinsic  galaxy shape,  which  means that  the measured  ellipticity
$\bm{\epsilon}^{({\rm obs})}$
\footnote{We use the complex ellipticity
 $\bm{\epsilon}=\epsilon e^{2i \psi}$, where $\epsilon = (1-r)/(1+r)$,
 and $r=b/a$ is the ratio between minor and major axes.}
of a galaxy can be expressed as \citep{2001PhR...340..291B, 2006glsw.conf.....M},
\begin{equation}
  \bm{\epsilon}^{({\rm obs})} = \bm{g} + \bm{\epsilon}^{({\rm I})} + \bm{\epsilon}^{({\rm rnd})},
\end{equation}
where   $\bm{g}$  is  the   reduced  shear,   defined  as   $\bm{g}  =
\bm{\gamma}/(1-\kappa)$, and $\bm{\epsilon}^{({\rm rnd})}$
denotes the noise part in galaxy shape measurements, which 
is assumed to be uncorrelated with the other components.
In  the weak  lensing  regime, $\kappa$  is
small  and the  $\bm{\gamma} \simeq  \bm{g}$ assumption  is often
made.   The    intrinsic   shape    of a galaxy   is    described   as
$\bm{\epsilon}^{({\rm I})}$. Ideally,  if the intrinsic ellipticities
of galaxies are  isotropic, the lensing shear $\bm{g}$  can be derived
by averaging  over a  population of galaxies.  However, it is  not the
case  for  real  data, because of the presence of correlated  intrinsic
alignment of observed galaxies.

The  observed 2-point shear  correlation function  consists of the
following           contributions          \citep{2015PhR...558....1T,
  2016ApJ...824...77J, 2016MNRAS.456..207K},
\begin{equation} \label{eq:2pcf}
  \langle \bm{\epsilon}^{({\rm obs})}_{i} \bm{\epsilon}^{({\rm obs})}_{j} \rangle =
  \langle \bm{g}_{i} \bm{g}_{j} \rangle + \langle \bm{\epsilon}^{({\rm I})}_{i} \bm{g}_{j} \rangle + \langle \bm{\epsilon}^{({\rm I})}_{i} \bm{\epsilon}^{({\rm I})}_{j} \rangle,
\end{equation}
where  we assume  that the  two observed  galaxies are located 
at  the redshifts $z_{i}$ and  $z_{j}$ (with $z_{i} \leqslant z_{j}$), respectively.
The  first term, $\langle  \bm{g}_{i} \bm{g}_{j}  \rangle$, represents
the shear-shear correlation, GG, which is the weak lensing 
signal we want to extract. The   correlation  $\langle
\bm{\epsilon}^{({\rm I})}_{i}  \bm{g}_{j} \rangle$  , often  named as
GI,  is  the cross  term  between  gravitational  shear and  intrinsic
ellipticity.  This  correlation  comes  from  the  fact  that  the shape of 
a distant galaxy `$j$'  is lensed  by the  foreground gravitational
potential, in which  galaxy `$i$' is intrinsically aligned with
the  underlying tidal  field  \citep{2004PhRvD..70f3526H}. Since
nearby galaxies are affected  by the same environment,  
the     intrinsic-intrinsic      correlation
$\langle \bm{\epsilon}^{({\rm I})}_{i}           \bm{\epsilon}^{({\rm I})}_{j}
\rangle$,  often  referred  as  II term, may be non-zero.
Both  the II  and  GI correlations  can contaminate  our measurements  of 
cosmic  shear, and are important to quantify, particularly  in accurate 
shear measurements expected from future large  lensing surveys
\citep{2016MNRAS.456..207K}.

In order to model the II and GI parts in the measurements,
\citet{2004PhRvD..70f3526H}, \citet{2007NJPh....9..444B} and \citet{2011A&A...527A..26J}
developed a non-linear intrinsic alignment model based on the work of \citet{2001MNRAS.323..713C}.
In this model, the power spectra of the II and GI contributions are related
to the non-linear matter power spectrum as $P_{\rm II}(k, z) = f^{2}(z) P_{\delta}(k, z)$
and $P_{\rm GI}(k, z) = f(z) P_{\delta}(k, z)$, respectively. Here
the modification factor $f(z)$ is defined as
\begin{equation}
  f(z) = -A_{\rm IA} C_{1} \rho_{\rm c} \frac{\Omega_{\rm m}}{D(z)}
         \left( \frac{1+z}{1+z_{0}}\right)^{\eta}
         \left( \frac{L}{L_{0}}\right)^{\beta},
\label{equ:IA}
\end{equation}
where $A_{\rm IA}$ is a free parameter;
$C_{1}=5\times 10^{-14} h^{-2}{\rm M}_{\sun}^{-1}{\rm Mpc}^{3}$;
$\rho_{c}$ is the critical density at the present; and $D(z)$ is the linear
growth factor (normalized to unity at $z=0$). 
The free parameters $\eta$ and $\beta$ account for the dependence  
on redshift and luminosity around the pivot redshift $z_{0}$ and luminosity $L_{0}$.
Following the discussion of \citet{2017MNRAS.465.2033J} based on the CFHTLenS
data, we fix $\eta=0$ and $\beta=0$ in our model fitting.
These formulas are used in section 5.3.2 to fit the measurements of GI and II terms 
from our simulation. There we will see that the sign of $A_{\rm IA}$ 
actually depends on galaxy type.

\section{Numerical simulations}

In   this  section,   we  describe  the  $N$-body  simulations
(\ref{simulation:Nbody}),  the    spherical     ray-tracing     technique
(\ref{simulation:rt}),  and  the comparison between the measured power  spectra  from  our  lensing
simulations   and    those from the   non-linear model predictions.

\subsection{$N$-body Simulation} \label{simulation:Nbody}

\begin{table*}
  \centering
  \caption{The parameters of the two $N$-body simulations. Cosmological parameters are given as $\Omega_{{\rm m}}, \Omega_{\Lambda}$, $h$ and $\sigma_{8}$. $L_{{\rm box}}$ is the box-size, $m_{{\rm p}}$ is the particle mass, $l_{{\rm soft}}$ is the softening length.}
  \label{tab:Nbody}

  \begin{tabular}{cccccccc}
    \hline
    Simulation & $\Omega_{{\rm m}}$ & $\Omega_{\Lambda}$ & $h$ &$\sigma_{8}$ & $L_{{\rm box}}/h^{-1}{\rm Mpc}$ & $m_{{\rm p}}/(10^{10}h^{-1}{\rm M}_{\sun})$ & $l_{{\rm soft}}/h^{-1}{\rm kpc}$  \\
    \hline
    PS-I & 0.260 & 0.740 & 0.710 & 0.80 & 1000 & 0.249 & 7.0 \\
    L500 & 0.282 & 0.718 & 0.697 & 0.82 & 500  & 0.034 & 3.5 \\
    \hline
  \end{tabular}
\end{table*}

We  use   two  sets  of   different  $N$-body cosmological  simulations. 
The  first  one  is a  part  of the  ELUCID project        
\citep{2014ApJ...794...94W,        2016ApJ...831..164W,
  2016RAA....16..130L,   2017arXiv170403675T}, which  is  run   with
$3072^{3}$ dark matter particles in a cubic box with $L_{{\rm box}} =
500 h^{-1} {\rm Mpc}$  on each side.  This simulation  is referred as
L500 in the following.   The  cosmological parameters  of  L500  are  from the  WMAP9
cosmology \citep{2013ApJS..208...19H}.   The second simulation  is the
Pangu  simulation  (PS-I), performed  by  the Computational  Cosmology
Consortium  of China \citep{2012ApJ...761..151L},  which has  the same
number of particles as L500, but with a  box size of $1000 h^{-1}{\rm Mpc}$ on each side.  The
cosmological  parameters   of  PS-I  are  from   the WMAP7
\citep{2011ApJS..192...18K}.   In Table.~\ref{tab:Nbody}, we  list the
parameters of  the two $N$-body  simulations.  Both simulations
were run using the GADGET-2 code \citep{2005MNRAS.364.1105S}.

With  its higher  mass resolution,  the  L500 simulation  is used  to
generate     galaxies     from     a      semi-analytical      model
\citep{2016MNRAS.458..366L}.   This model is  based on  the L-Galaxies
model        developed        by        the        Munich        group
\citep[e.g.,][]{2013MNRAS.428.1351G} (see Section 4.1 for more details).
We do  not produce  mock galaxies using the PS-I simulation due to its lower mass resolution, but use it
as a reference to check our calculation of the convergence power spectrum on large scales.

\subsection{Spherical Ray-tracing Simulation} \label{simulation:rt}

To perform a ray-tracing simulation with full-sky coverage,
we  follow  the  multi-plane  algorithm
developed by \citet{2008ApJ...682....1D}, \citet{2009A&A...497..335T},
and  \citet{2013MNRAS.435..115B}.  In  order to  control the
residual in  the solution of  the lensing potential, we  implement
an  iterative   spherical  harmonic  analysis scheme, which is different from the
multi-grid method adopted by \citet{2013MNRAS.435..115B}. 
In the following, we briefly  summarize  the  main procedures; 
more   details   can  be   found  in Appendix~\ref{appendix:rt}.

To trace  the trajectory of  a light beam,  we first employ
the   $N$-body   simulations   to   build   light-cone   to   redshift
$z_{{\rm max}}  \sim  2.0$.  In  practice the  simulation  boxes  are
divided into sets of  small cubic boxes with $\sim 100 h^{-1}{\rm Mpc}$ on
each  side. These  cell  boxes  are appropriately piled together so as
to cover  the past light-cone from $z=0$ to $z=z_{{\rm max}}$.
For both L500 and PS-I  described above, a full-sky 
light-cone can be constructed in this manner.
Note that the size of our simulation box is relatively small 
compared with the comoving distance to redshift $z_{\rm s} = 2.0$ 
and periodic effects will show up at several specific directions, 
especially along the box axes, but the effects disappear very quickly apart from these directions.

Then each light-cone is divided into a set of spherical shells with a
thickness of $50 h^{-1} \rm Mpc$ centered at the observer, and the dark  matter distribution in
the corresponding  shells  is  projected into  pixels defined by
the    HEALPix    \footnote{\url{healpix.jpl.nasa.gov}}   tessellation
\citep{2005ApJ...622..759G,  2007MNRAS.381..865C}. The
HEALPix resolution parameter is set to $N_{{\rm side}} = 8192$, which
gives an angular resolution  of $\sim 0.43$ arcmin.\footnote{Given the
  HEALPix resolution $N_{{\rm side}}$,  one can calculate the
  pixel  scale   by  ${\rm d}\theta  =  \sqrt{4  \pi   /  (12  \times
    N^2_{{\rm side}})}$}   The projected  surface   mass   densities  are
calculated   for   each    shell   using   the SPH    algorithm
\citep{2005ApJ...635..795L, 2010ARA&A..48..391S}. We  use the nearest
64  particles to define  the kernel size, but keep  the smoothing
length larger than  two HEALPix  cells in high-density regions. The
lensing potential for the  $n$-th shell, $\phi^{(n)}_{\ell m}$, is then
obtained,  using the  Poisson equation,  from the  mass  density shell
after applying an  iterative spherical harmonic transformations (refer
to as HEALPix predefined functions),
\begin{equation}
  -\ell(\ell+1)\phi^{(n)}_{\ell m} = 2\kappa^{(n)}_{\ell m}.
\end{equation}
To  perform multi-sphere ray-tracing  simulations, we set
the initial  positions of ray-beams at the centers  of the HEALPix
cells,  and  propagate light  rays  from the observer  to a desired
redshift applying deflection angle,
\begin{equation}\label{equ:alpha1}
  \alpha^{(n)}_{\ell m} = -\sqrt{\ell(\ell+1)}\phi^{(n)}_{\ell m}.
\end{equation}

\begin{figure}
  \centering
  \includegraphics[width=0.98\columnwidth]{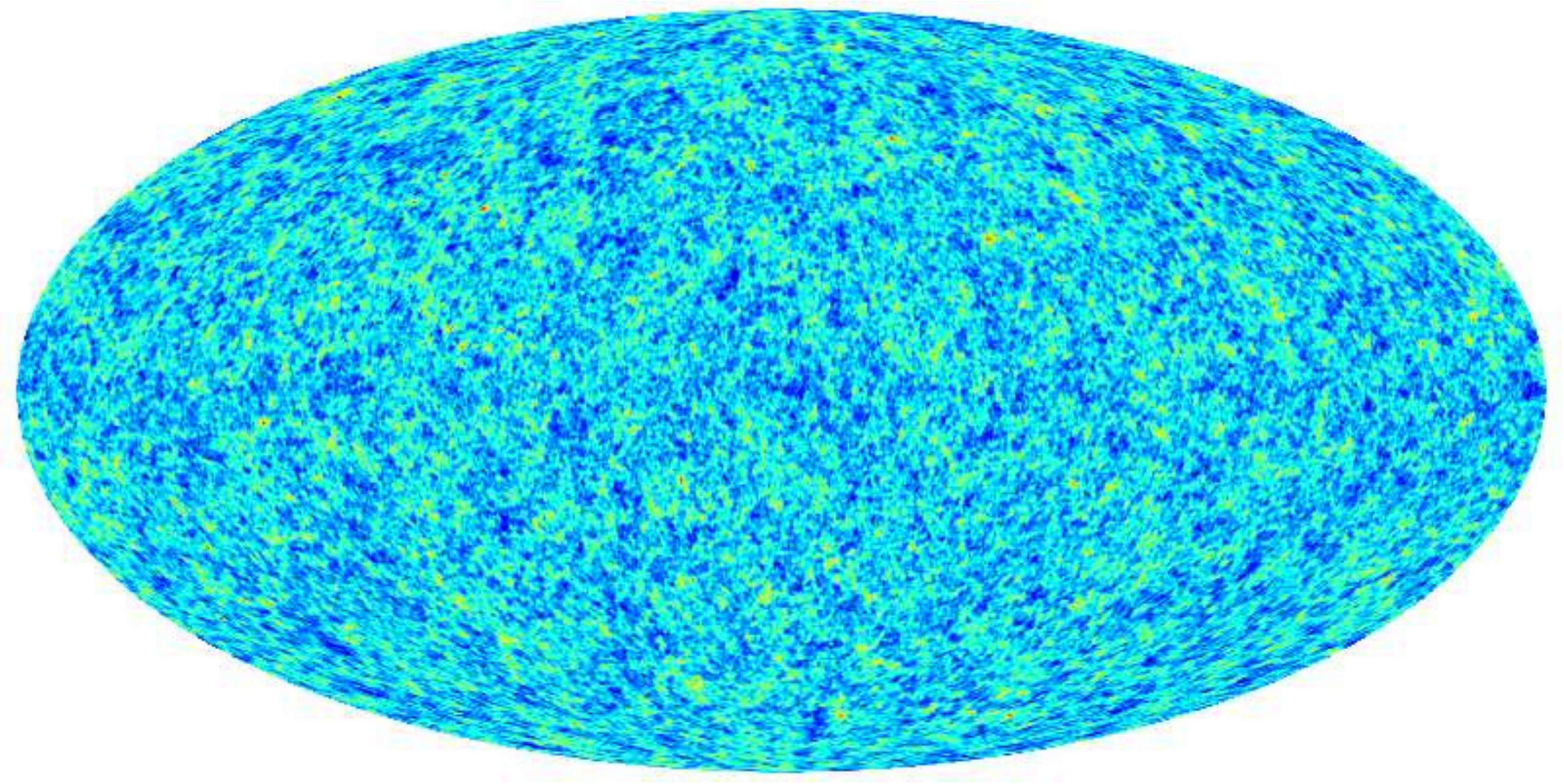}
  \caption{One realization of the convergence map from PS-I light-cone for sources at $z_{\rm s} = 1.0$.}
  \label{fig:kmap}
\end{figure}

\begin{figure*}
  \centering
  \includegraphics[width=0.98\columnwidth]{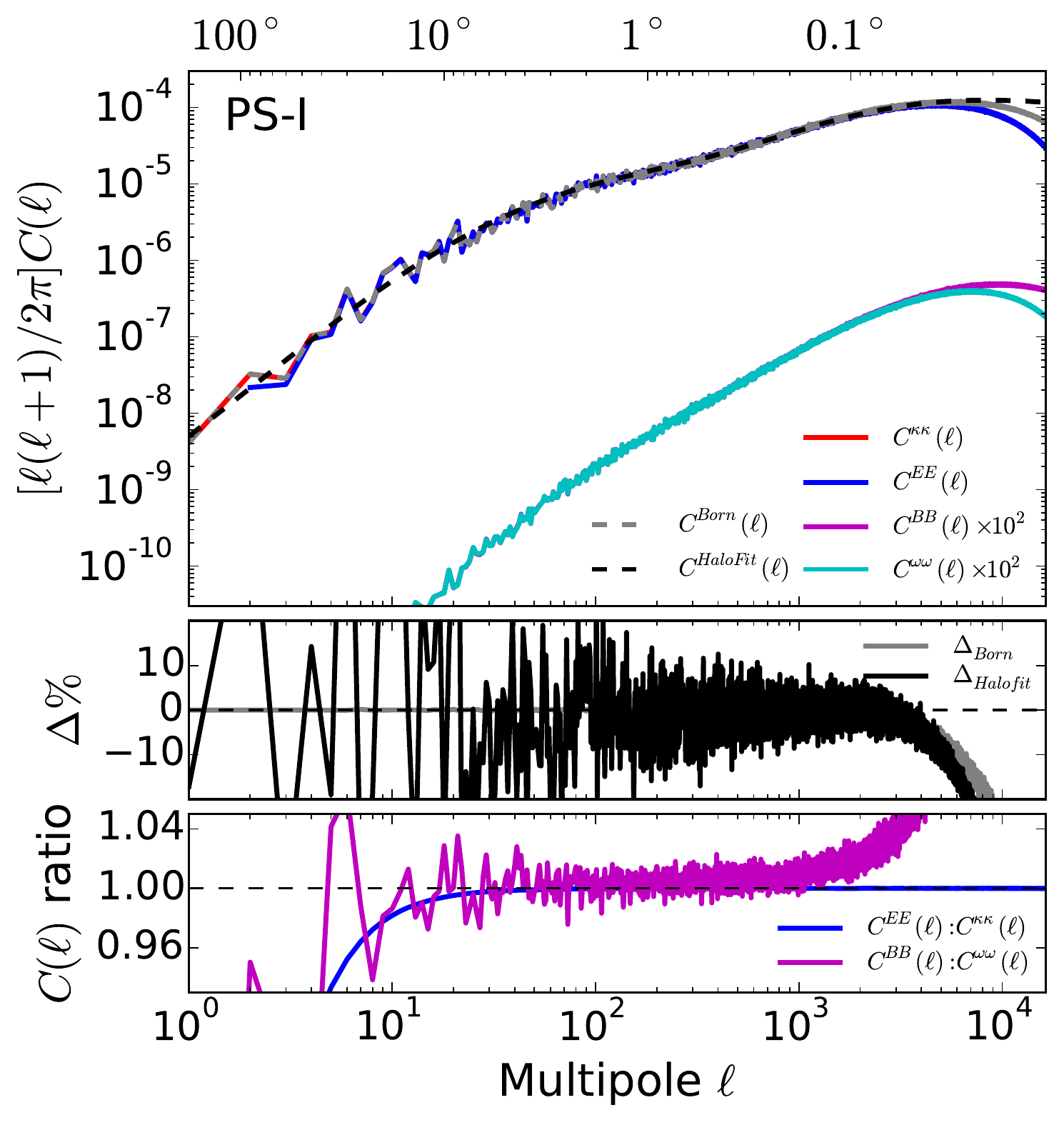}
  \quad
  \includegraphics[width=0.98\columnwidth]{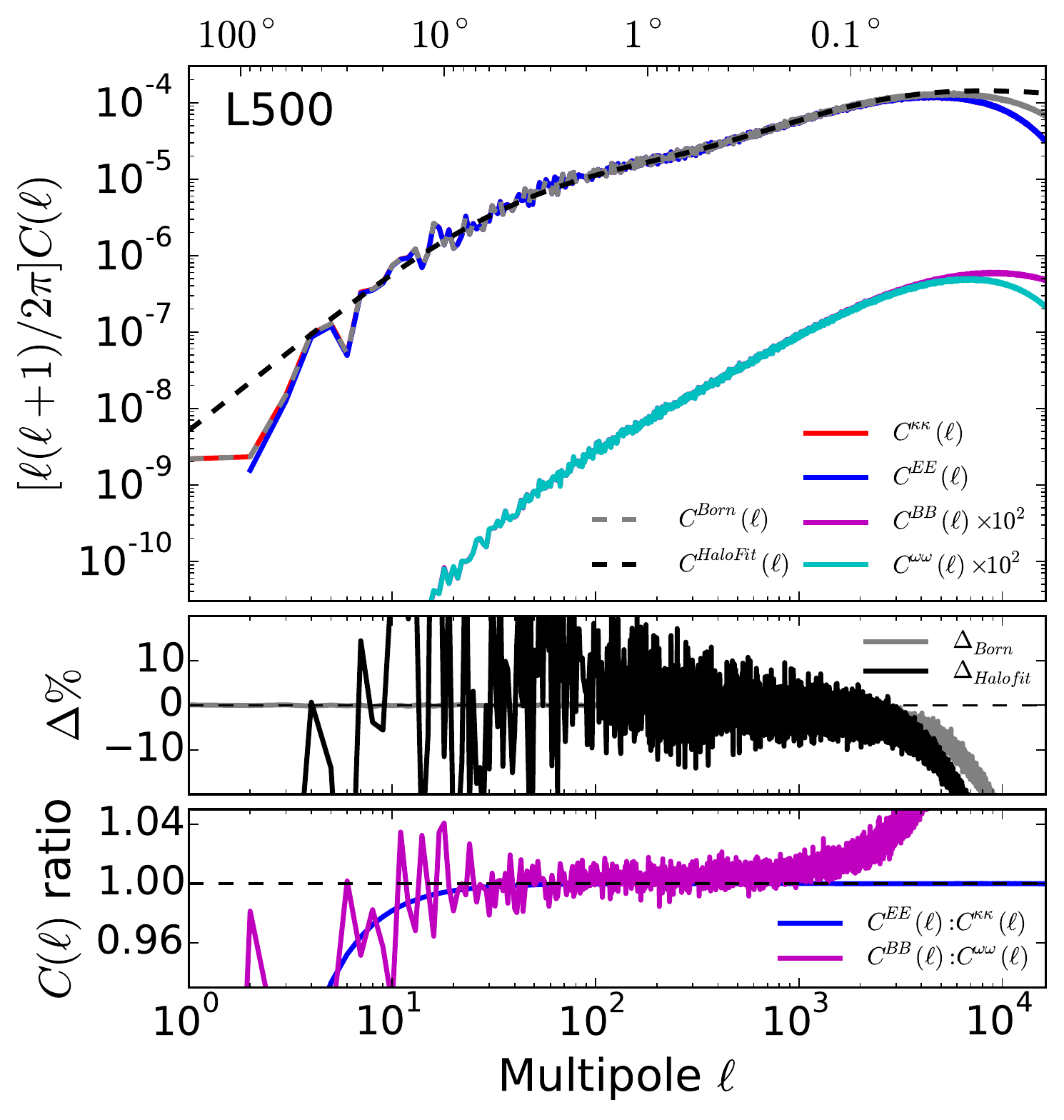}
  \caption{Comparison between the power spectrum from ray tracing
  with the model predictions. The top panels show the angular power
  spectrum of convergence (red solid line), shear E-mode (blue) and B-mode (magenta),
  and rotation (cyan) for sources at $z_{s} = 1$ from the PS-I (left panel)
  and L500 (right panel). The measured convergence power
  from the Born approximation (grey dashed line) and revised Halofit model
  \citep{2012ApJ...761..152T} predictions (black dashed line) are also shown
  for comparison. The relative deviations of the measured convergence
  statistics and from predictions are presented in the middle panels.
  Lower panels show the ratios of the E- and B-modes with to
  the convergence and rotation modes, respectively.}
  \label{fig:PS1}
\end{figure*}

From  our  lensing simulation,  we evaluate  the
distortion matrix  $\mathcal{A}$ on  each lensing shell  and construct
the full-sky map of the convergence and lensing shear.   As  an   illustration,
Fig.~\ref{fig:kmap} shows one  realization of our simulated full-sky
convergence  map, $\kappa$,  for  sources  at redshift $z_{s}  =  1.0$.
In Fig.~\ref{fig:PS1},  we  show the  power spectra measured from
PS-I  (left panels) and L500 simulation (right panels).  The  top panels  show  the angular  power
spectra of the convergence (red solid line), the shear E-mode (blue) and B-mode (magenta), and the rotation mode (cyan).
We   also  show   the   prediction  from the  Born   approximation
\citep{2002ApJ...574...19C} by stacking density field along the line-of-sight
in our mock light-cone as   the grey dashed line
and the theoretical prediction from the     revised    non-linear    Halofit
\citep{2012ApJ...761..152T, 2014ascl.soft02032P} as the black dashed line.
The  middle  panels  of Fig.~\ref{fig:PS1}  show  the
relative  deviations  between  convergence powers measured from  our
ray-tracing simulation and the theoretical predictions, like that
the relative deviation between the ray-tracing simulation and Halofit model is defined by
$\Delta_{\rm Halofit} = \left[ C^{\kappa\kappa}(\ell) - C^{\rm Halofit}(\ell) \right]/ C^{\rm Halofit}(\ell)$.

The measured  convergence  power from our ray-tracing simulation
agrees well with the theoretical prediction of Halofit, and the 
relative error is less than 10 per cent at $\ell \lesssim 4000$. 
At large scales,  $\ell \lesssim 10$, the 
power from the PS-I  simulation is in better agreement with theoretical
predictions than the L500 simulation, as is expected from the fact that
PS-I has a larger box to represent the matter power on large scales. 
The prediction from Born  approximation is closer  to 
that of Halofit  at  small  scales than  our  ray-tracing
simulation, because both Born  approximation and Halofit
are  based on the first-order  approximations.  This  good
agreement, within 10 per cent for  $\ell \lesssim 6000$, roughly
on scales larger than the smoothing scale in our simulation, indicates 
that  the revised Halofit model  \citep{2012ApJ...761..152T} provides a  
good approximation to the non-linear matter power spectrum.  
In addition,  since  the power  spectra  from our  Born
approximation  and full-sky  ray-tracing simulation  are both 
based  on the same convergence $\kappa$ maps,  the difference between 
the two is not caused by the smoothing effect. 
The grey lines in the middle panels of Fig.~\ref{fig:PS1} show that
Born approximation can cause a deviation of more than 10 per cent 
for $\ell > 6000$.

The bottom panels  show the ratio of shear  E-mode and B-mode  power spectra
relative to  the measured  convergence and rotation mode spectra, respectively.
As shown in  Section~\ref{sec:introPS},  at high-$\ell$  (small
scales), one has $C^{EE}(\ell) = C^{\kappa\kappa}(\ell)$ from
the full-sky weak lensing. Our corresponding measurements on small scales
are   indeed   in  consistent   with   this expectation.
 At   large   scales   (low-$\ell$), we   must   take  account
of the extra factor of $(\ell-1)(\ell+2)/\ell/(\ell+1)$ to explain the
difference   between  the   power   spectrum  of   the shear  E-mode   
and convergence spectrum. 
As discussed in \citet{2013MNRAS.435..115B}, we 
also measure the power spectra of the B-mode and the rotation mode
from our lensing simulation. We find that the  B-mode power 
is  effectively suppressed relative to the E-mode by more than four
orders of magnitude. Moreover, the power ratio between the B-mode and the 
rotation mode shows that the extra numerical B-mode in our simulation
is negligible, and the accuracy of our shear map is only limited by the smoothing length at small scales.
We  refer the reader to Fig.~\ref{fig:dcl2} in Appendix A for more
details. Compared  with the \citet{2013MNRAS.435..115B} results, the predictions 
of our simulations for the convergence and shear power spectra are more 
accurate extending to higher $\ell$.

\section{GALAXY MODELING}\label{sec:GM}

One  important merit  of our  work is  to include  model  galaxies in 
the N-body simulations and to predict shear correlation functions  
that can be compared directly to observations.  In this section,  we 
describe how we model the  physical properties  of galaxies and show  
the intrinsic alignment of the model galaxies.

\subsection{Semi-analytic Models}

\begin{figure}
  \centering
  \includegraphics[width=0.98\columnwidth]{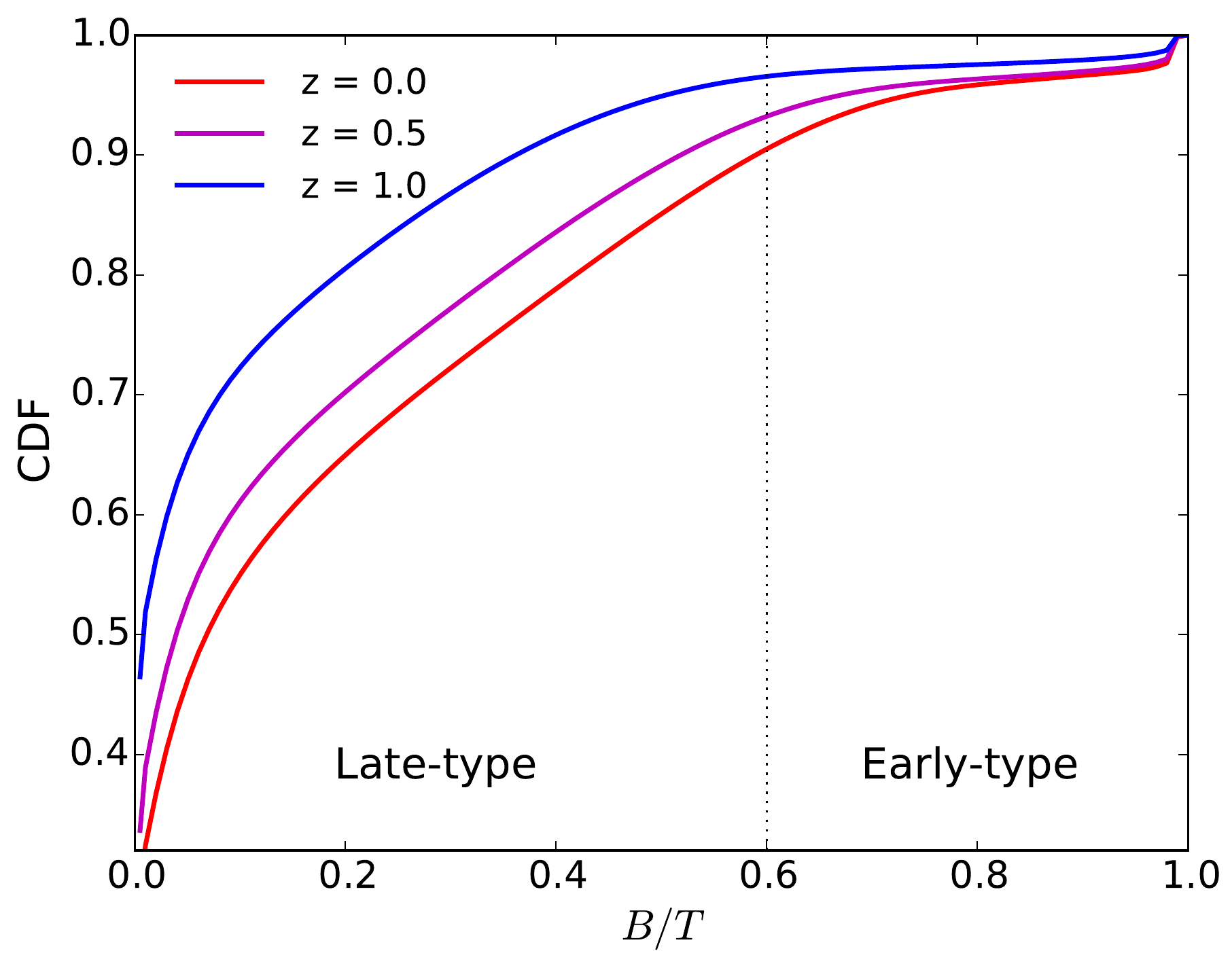}
  \caption{Cumulative probability distributions of the
  bulge-to-total mass ratio of model galaxies at three different redshifts.}
  \label{fig:btCDF}
\end{figure}

The model galaxies are produced using the semi-analytical model of
\citet{2016MNRAS.458..366L} which is based on the \citet{2013MNRAS.428.1351G} model,
one version of the Munich Semi-analytical model which is called L-Galaxies. 
In a semi-analytical model, galaxy population is assigned to dark matter haloes 
on the basis of simple assumptions of many physical processes. As a first step in our implementation 
of L-Galaxies, dark matter haloes are identified in the $N$-body simulation using the standard Friends-of-Friends
(FOF) algorithm. Only haloes that contain at least 20 particles are used.  
The subhaloes within each FOF halo are identified with the SUBFIND 
algorithm \citep{2001MNRAS.328..726S, 2005Natur.435..629S}.
Merger trees of these dark matter (sub)haloes can be constructed by linking 
progenitors of a halo in different snapshots. 
Galaxies are assumed to form at the centers of the dark matter haloes 
according to analytical prescriptions of the relevant physical processes, 
such as gas cooling, star formation, supernova and black hole
feedback. For the details of L-Galaxies, we refer the reader to \citet{2013MNRAS.428.1351G}.
\citet{2016MNRAS.458..366L} improved the prescription for low-mass galaxies, 
especially satellite galaxies, by including additional physics about cold gas 
stripping and an analytical modeling of orphan galaxies. In this model, the stellar mass
function, HI and H$_{2}$ mass functions are tuned to match the observational data
\citep{2003ApJ...582..659K, 2005MNRAS.359L..30Z, 2008MNRAS.388..945B, 2009MNRAS.398.2177L}.
The fraction and spatial distributions of central versus satellite galaxies 
are reproduced roughly correctly by the model, as shown in \citet{2016MNRAS.458..366L}
and \citet{2013MNRAS.428.1351G}.

To describe the morphology of a galaxy, we use the ratio between
the bulge and the total mass ($B/T$), which can be predicted from
the semi-analytical model, 
as the classification of `early-type' and `late-type' galaxies \citep{2009MNRAS.396.1972P}.
Following \citet{2013MNRAS.431..477J}, we adopt $B/T = 0.6$ to classify
the model galaxies into early  or late types. Fig.~\ref{fig:btCDF}
shows  the cumulative probability distribution of
$B/T$ of our simulated  galaxies  at three different redshifts.  
More  than  $80\%$  of all galaxies are  late-types, and 
the fraction is slightly  higher at high redshifts. This
fraction of  early/late-types is consistent with that found in other studies
\citep[e.g.,][]{2013MNRAS.428.1351G}. However, compared 
to hydro-dynamical simulations, semi-analytical models are less 
powerful in predicting the shapes of galaxies. To proceed,
we have to assign shapes to galaxies and their images with 
some simplified prescriptions, as described below.

\subsection{Galaxy Shape Measurement}\label{sec:galaxyshape}

A common assumption is that  the shape  of an elliptical galaxy roughly 
follows  that of its host dark matter halo, while the rotation axis of a spiral galaxy 
is determined by the spin of its halo  \citep[e.g.,][]{2007MNRAS.378.1531K,  2009ApJ...694..214O,
  2010ApJ...709.1321A}.                              \citet[][hereafter
  'J13']{2013MNRAS.436..819J} used this assumption  and studied
the   alignment   of   galaxies   from   the   Millennium   Simulation
\citep{2005Natur.435..629S}.  Here we follow J13  to assign 
shapes to model galaxies. 

To assign  a shape to a model galaxy  using the mass distribution  of its dark
matter  halo, we need to  distinguish between  central  and satellite
galaxies.  A central galaxy  is assumed to be located at  the center of  a 
dark matter halo  and its shape may be related to that of the host halo.
The shape of a dark matter halo  is usually defined  using the inertia
tensor $I_{ij}$ \citep{2005ApJ...627..647B},
\begin{equation}
  I_{ij}= \sum^{N_{\rm p}}_{n=1}{m_{p} x_{i,n}, x_{j,n}},
\end{equation}
where $N_{\rm p}$ denotes the particle number of the FOF halo, and
${\bf x}_{n}$  is the position  of the  $n$-th particle  with respect  to the
center of the halo. By diagonalizing the inertia tensor $\bm{I}$, one
can get  the eigenvalues $\lambda_{1}  \leqslant \lambda_{2} \leqslant
\lambda_{3}$  and  the corresponding  eigenvectors  that define 
a  triaxial ellipsoid and its orientation. It has been argued that 
ta  minimum number  of  $N_{\rm p}=300$ is  needed  to ensure  an
accurate  measurement  of  the halo   shape 
\citep{2002MNRAS.335L..89J, 2007MNRAS.376..215B}. 
Constrained by the  resolution  of  our simulation,  we  have to reduce  
the  number  limit  to $100$.  As shown in J13, a minimum number of $100$  
can lead to $\sim10\%$ deviation in the  axis ratio and $10^{\circ}$ 
deviation in the orientation angle. Since the magnitude limitation, the 
corresponding haloes always have more than 100 particles in our mock catalogs
for DLS and KiDS-450.
Once the three-dimensional shapes  and orientations of galaxies are obtained, 
we project them into the sky  to     obtain    the     projected    ellipses
\citep{1983Ap&SS..92..335G,       1985MNRAS.212..767B}.       
Details about how to make the projections      can     be      found      in
appendix~\ref{appendix:proj}.

In  addition  to the  above  model in which perfect  alignments are assumed 
between elliptical galaxies and their host halos, J13 also considered a mis-alignment 
model in which the major axis  of the central  elliptical galaxy  is mis-aligned 
with that  of the halo, with the mis-alignment  angle obeying a gaussian 
distribution with zero mean and a dispersion of $35^{\circ}$.  This is motived
by the finding  that such a mis-alignment  is needed to explain the alignment
between luminous red  galaxies on large scales 
\citep[e.g.,][]{    2007MNRAS.378.1531K,
  2009RAA.....9...41F, 2009ApJ...694..214O, 2013ApJ...768...20L}.  
We will come back in Section~\ref{sec:GIIIterm} to discuss the effect of such mis-alignment     
on     shear    correlation    functions.

For a central late-type galaxy, defined by $B/T < 0.6$, J13 assigned its
shape according to the angular momentum vector of the host halo,
\begin{equation}
  \bm{L} = \sum^{N_{\rm p}}_{n=1}{m_{p} \bm{x}_{n} \times \bm{v}_{n}},
\end{equation}
where $\bm{v}_{n}$  is the velocity  of the $n$-th halo particle relative  
to the halo center. The angular momentum  $\bm{L}$ defines a circular disc in
the  halo, and the complex  ellipticity is obtained  by
projecting the disc along the line  of sight.  As described in J13, the apparent 
axis ratio of the projected ellipse is,
\begin{equation}
  r = \frac{|L_{{\rm los}}|}{|\bm{L}|} + r_{d}\sqrt{1-\frac{L_{{\rm los}}^2}{|\bm{L}|^2}},
\end{equation}
where $r_{d}$  is the ratio  between the disc thickness  and diameter, 
and we  set $r_{d}=0.25$ following J13; $L_{{\rm los}}$ is
the component of $\bm{L}$ along the line of sight. The
ellipticity of the mocked galaxy is then given by
\begin{equation}
  \epsilon = (1-r)/(1+r).
\end{equation}

For a satellite galaxy, on the other hand,  the original dark matter halo 
associated with it may have suffered strong mass loss after it is accreted
into a big halo, depending on the infall time and orbit \citep{2000MNRAS.319..168C}.
It is thus unclear how the shapes of satellites are connected to those 
of the dark matter subhaloes associated with them.  Some earlier 
investigations \citep[e.g.,][]{2008ApJ...672..825P} have shown  that the tidal torque
of the host halo can induce a correlation between subhalo orientation
and its direction to the host center.  
Therefore, J13 assigned the shape of an early-type satellite galaxy by randomly
choosing the three-dimensional axis ratios from a halo sample with more than 300 
particles and then made its major axis point to the central  galaxy.
For a late-type satellite,  its  spin is  assumed to 
be perpendicular  to  the  line connecting  the satellite to the  central galaxy, and 
the ellipticity is obtained by projecting  the disk onto the  sky, as done for  the 
central spiral galaxies.

The  above  assumptions for  the shapes and orientations  of  
satellite galaxies  are clearly too idealistic. In fact, while 
orbiting in their host halos, satellite galaxies may have 
their  radial  alignments scrambled.   Observationally,  
measurements of  the shapes of faint  satellite galaxies are
difficult and  sensitive to  the methods used  to derive  galaxy shapes
\citep[e.g.,][]{2011ApJ...740...39H}. Currently, there   is  no   consensus
on  the alignments  of satellites.  Some studies have   reported   detection   
of  radial   alignment   of   satellites
\citep{2005ApJ...627L..21P,  2006ApJ...644L..25A, 2007ApJ...662L..71F,
  2014MNRAS.445..726C,    2015MNRAS.450.2195S,   2016MNRAS.463..222H},
while     others    have     not     found     such    an     alignment
\citep{2009arXiv0903.2264S,  2011ApJ...740...39H, 2013MNRAS.433.2727S,
  2015A&A...575A..48S}.   Because of this  uncertainty, we  also consider 
 a  simple case in  which the orientations  of satellites,  regardless  of 
 their types,  are random distributed in their  host haloes.  We will show
in Section 5.3 that satellite alignment has a stronger effect on the
shear correlation on smaller scales.

\subsection{Intrinsic Shape Correlations of Mock Galaxies}

\begin{figure}
  \centering
  \includegraphics[width=0.98\columnwidth]{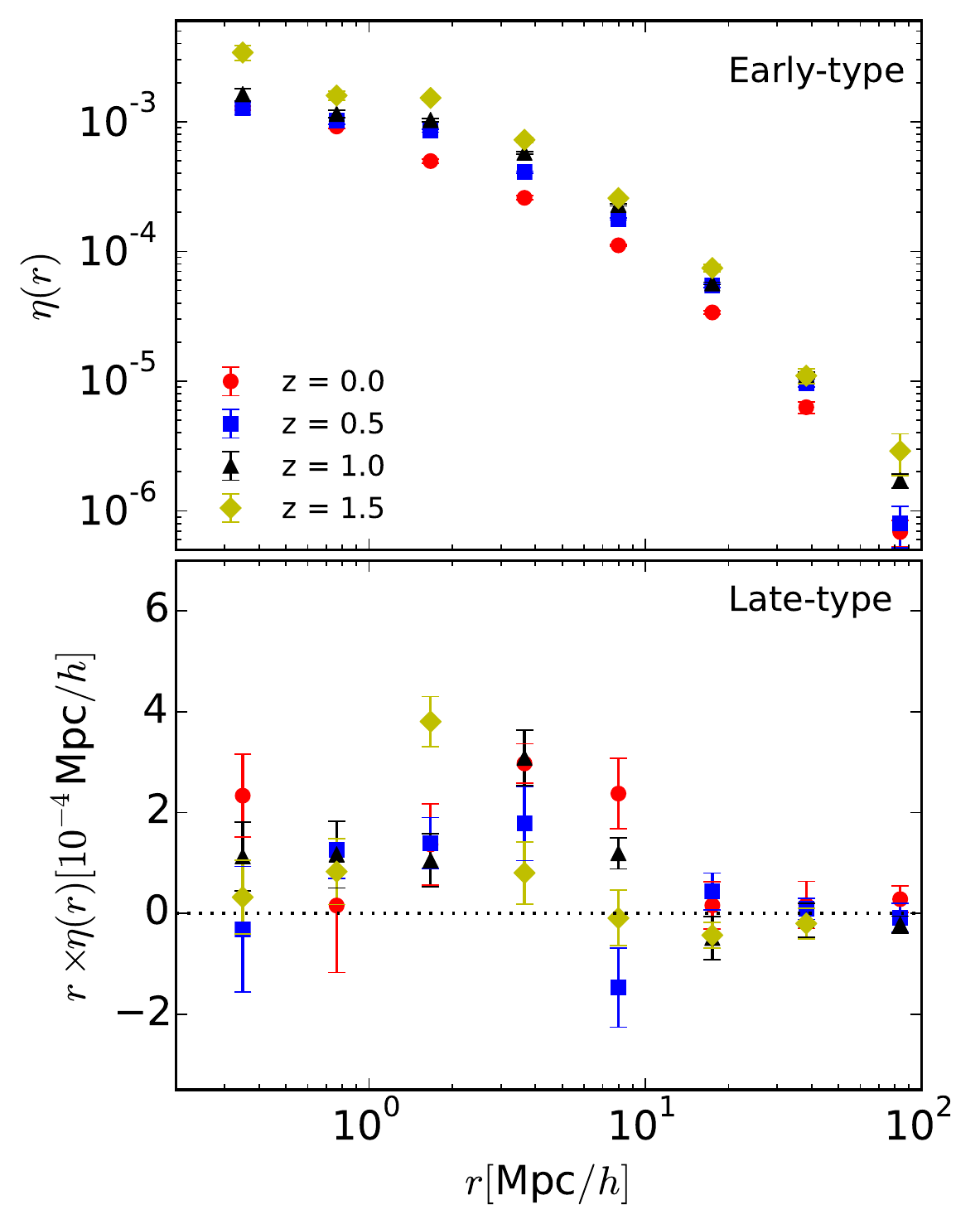}
  \caption{The intrinsic ellipticity correlation, $\eta(r)$,
  for early-type (upper panel) and late-type (lower panel) central
  galaxies. Only central galaxies within haloes of mass
  $m_{{\rm halo}} \geq 3.4\times10^{10} {\rm M}_{\sun}/h$ (i.e.
  with 100 particles or more) are used. 
  Early-type galaxies show
  significant shape alignments and redshift dependence, while for
  late-types the correlations of their intrinsic ellipticities
  are very weak. Here errors are estimated using the jackknife method.}
  \label{fig:eta0}
\end{figure}

\begin{figure}
  \centering
  \includegraphics[width=0.98\columnwidth]{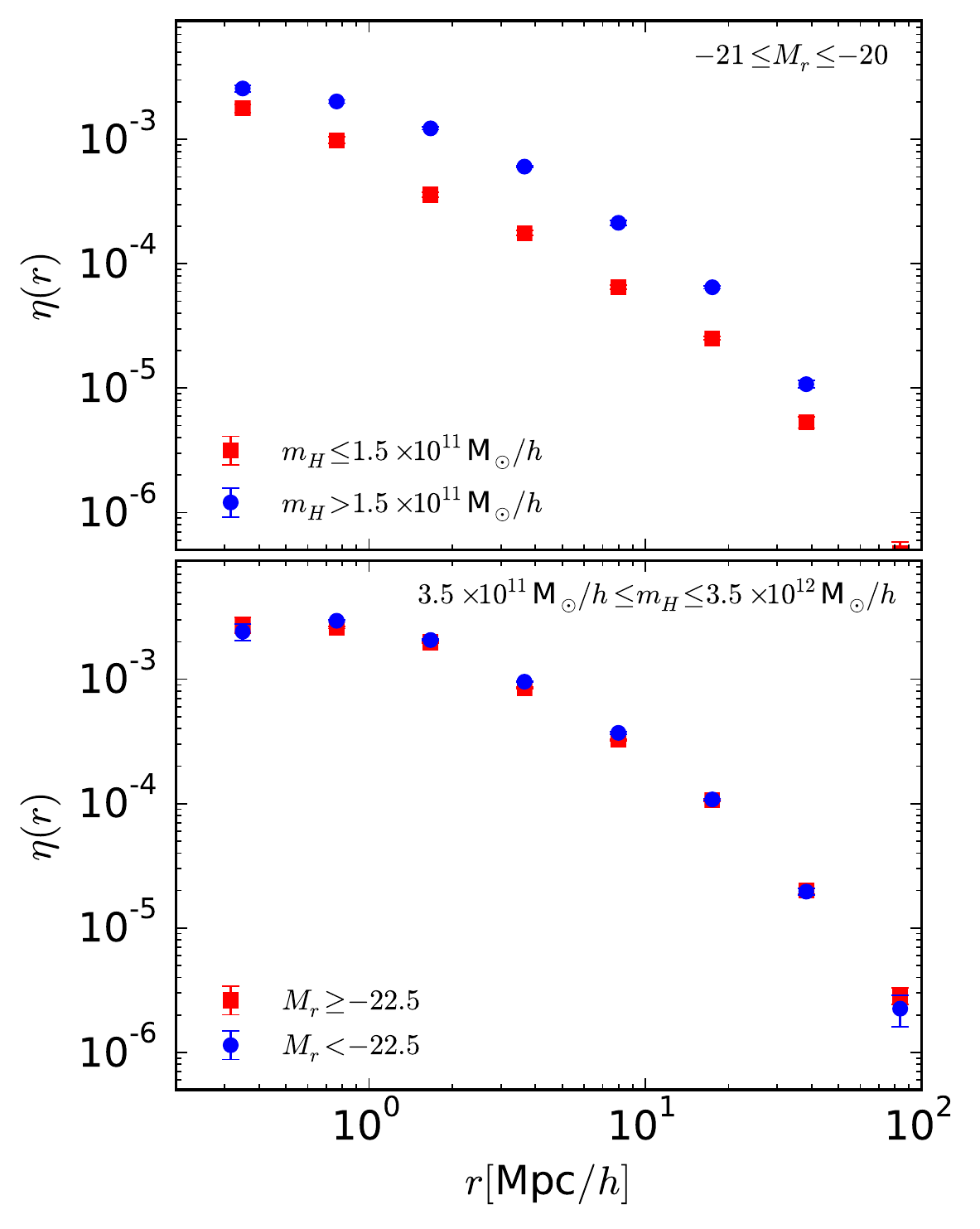}
  \caption{Correlation function $\eta(r)$ of early-type galaxies at $z = 1.0$.
  The top panel shows the halo mass dependence of the correlation for galaxies
  in a given luminosity bin.
  The bottom panel shows the correlation of galaxies in a given small range of
  halo mass but with different luminosities. The figure shows that {the intrinsic ellipticity}
  correlation is mainly determined by halo mass.}
  \label{fig:eta1}
\end{figure}

The  intrinsic shape  correlation  function of  galaxies, $\eta(r)$, 
is defined as 
\begin{equation}\label{equ:eta}
  \eta(r) = \langle \epsilon_{t}(\bm{x}) \epsilon_{t}(\bm{x}+\bm{r}) + \epsilon_{\times}(\bm{x}) \epsilon_{\times}(\bm{x}+\bm{r}) \rangle_{\bm{x}},
\end{equation}
where $r$ is the three-dimensional comoving separation between two galaxies
\citep[e.g.,][]{2006MNRAS.371..750H}. The quantities 
$\epsilon_{t}$ and $\epsilon_{\times}$ are the tangential and cross components of
the galaxy ellipticity: 
\begin{equation}
\epsilon_{t}+{\rm i}\epsilon_{\times} = -\bm{\epsilon} {\rm e}^{-2{\rm i}\varphi}
\end{equation}
where $\varphi$ is the angle between the separation vector
of a given galaxy pair and the horizontal axis \citep{2001PhR...340..291B}.

Following \citet{2013MNRAS.436..819J}, we first measure the
intrinsic shape correlation of galaxies in our simulation by projecting
semi-analytic galaxies along the line of sight parallel to the edges of
the simulation box. In order to estimate the error bars in our measurements,
we divide our simulation box into eight equal-sized cubic sub-boxes of
$250\,{\rm Mpc}/h$, and use the Jackknife method to estimate the errors.
Fig.~\ref{fig:eta0}  shows   the  redshift  dependence  of  the
correlation  function $\eta(r)$ for  early-type (upper)  and late-type
(lower) central  galaxies in our simulation. It is seen that early-type
galaxies have a  strong correlation and the correlation is stronger at 
higher redshifts. In contrast, late-type galaxies do not show any 
significant correlation of their projected ellipticities, although 
some weak positive correlation signals  can be seen  at small  scales, 
$r \lesssim  8 {\rm Mpc}/h$. The  weak/null correlation for spiral  galaxies in
our  simulation   is  consistent   with  the  non-detection   in  both
observations \citep[e.g.,][]{2011MNRAS.410..844M} and in simulation results
\citep[e.g.,][]{2013MNRAS.436..819J}.

In Fig.~\ref{fig:eta1}  we further  investigate the dependence  of the
shear  correlation $\eta(r)$  on halo mass  and luminosity  for 
early-type galaxies  at  $z =  1.0$.   To separate  the two  dependencies, 
we select galaxies in a small 
ranges of luminosity and halo  mass, and  divide galaxies into  two
subsamples in  halo mass (the top panel) and in galaxy luminosity 
(the  bottom panel ).  It can be seen that there is no significant 
luminosity dependence for fixed halo mass,  but  a significant 
dependence on  halo  mass  is seen at a fixed luminosity.  
This dependence  of {the intrinsic ellipticity} correlation on mass and
luminosity in our simulation is similar to the results found in J13.

\section{COSMIC SHEAR AND COMPARISON WITH OBSERVATION}

\subsection{Shear Correlation Function}\label{sec:xipm}

Weak lensing  induces small correlated  distortions in observed
galaxy   shapes.   This   correlation  can   be  quantified   using  different
statistics.  Observationally,  the most  direct measurement  of  the 
lensing signal  is  the  two-point shear  correlation function. 
The shear-shear correlation  between galaxies at a given
separation $\vartheta$ is estimated as
\begin{equation}
  \xi_{tt}(\vartheta) = \frac{\sum_{i,j} w_{i}w_{j}\epsilon_{t,i}\epsilon_{t,j}}{\sum_{i,j}w_{i}w_{j}}
\end{equation}
and
\begin{equation}
  \xi_{\times\times}(\vartheta) = \frac{\sum_{i,j} w_{i}w_{j}\epsilon_{\times,i}\epsilon_{\times,j}}{\sum_{i,j}w_{i}w_{j}},
\end{equation}
where $w_{i}$ is the ellipticity weight of the $i$-th galaxy and $\vartheta$ is
the angular separation between the galaxy pair.

The two linear combinations of $\xi_{tt}$ and $\xi_{\times\times}$ 
that are also frequently used in the lensing analysis are 
\begin{equation}\label{equ:xipm}
  \xi_{\pm} = \xi_{tt} \pm \xi_{\times\times}.
\end{equation}
The convergence power spectrum $C^{\kappa\kappa}(\ell)$ can be related to
the estimator $\xi_{\pm}$ as 
\begin{equation}
  \xi_{\pm}(\vartheta) = \frac{1}{2\pi}\int^{\infty}_{0}{\rm d}\ell \ell J_{0,4}(\ell\vartheta) C^{\kappa\kappa}(\ell),
\end{equation}
where $J_{0,4}(\ell\vartheta)$ denotes the zeroth and fourth Bessel function for $\xi_{+}$ and $\xi_{-}$,
 respectively \citep{2002A&A...396....1S}.

\subsection{Tomographic Cosmic Shear}

Tomographic    measurement   is    capable    of    utilizing    the
redshift-dependence   of   cosmic  shear   signals   to  reveal   both
the cosmological structure  growth and the redshift-dependent  geometry in the
universe \citep{2003A&A...398...23K}, and
has  been widely used  in weak lensing  observations, such  as 
CFHTLens \citep{2013MNRAS.432.2433H}, DLS \citep{2016ApJ...824...77J},
and  KiDS-450  \citep{2017MNRAS.465.1454H}.   In  order  to
perform similar tomographic analysis  of cosmic  shear in  our mock
observation and compare the prediction with the KiDS and DLS observations,
we    employ   the    tomographic   redshift    bins   as    used   by
\citet{2017MNRAS.465.1454H}          for          KiDS-450         and
\citet{2016ApJ...824...77J} for DLS,  to mimic their measurements of cosmic
shears.

\begin{figure}
  \centering
  \includegraphics[width=0.98\columnwidth]{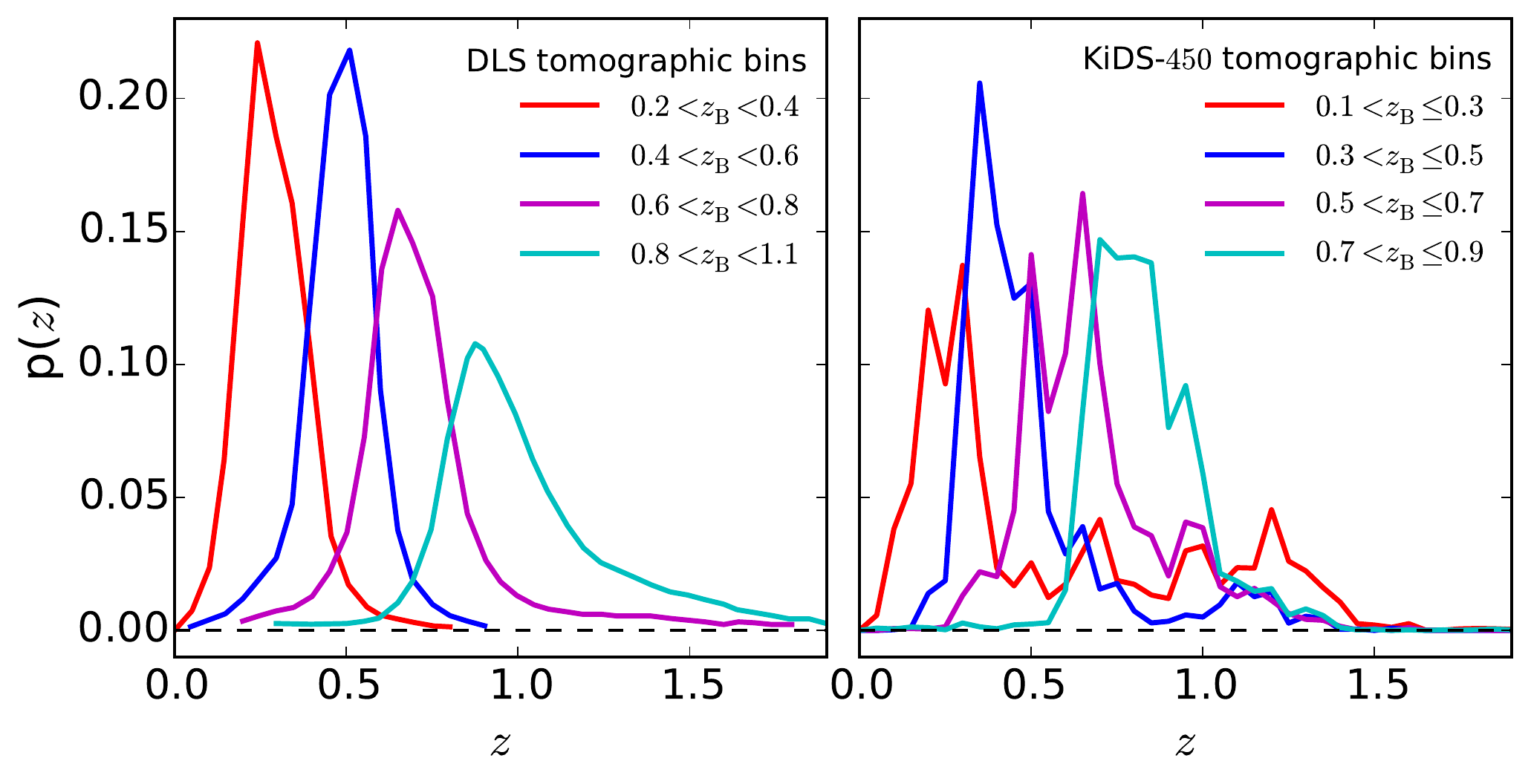}
  \caption{Redshift distributions of source galaxies
  in the two weak lensing surveys, DLS (left) and KiDS-450 (right).
  These redshift distributions are used in our light-cone to select
  source galaxies and used for the tomographic analyses.}
  \label{fig:KIDSandDLSphotoz}
\end{figure}

Fig.~\ref{fig:KIDSandDLSphotoz} shows the redshift distributions
of  the source galaxies  in the  two  surveys,
where $z_{\rm B}$ denotes the Bayesian point estimates of
the photo-$z$ \citep{2016ApJ...824...77J, 2017MNRAS.465.1454H}.
It  is seen  that the  two distributions  are   quite  different.   
In  KiDS-450   the  shape  of the distribution is  not regular, with more overlaps 
between different redshift  bins,  while the  distribution  for DLS is  more  regular  and
different bins are more clearly separated. 
For a consistent  comparison between model predictions  and observations, 
we  adopt their  redshift distributions for the  source  galaxies respectively, 
and  we truncate source galaxies at $z=2.0$.

To compare  with the survey results,  we first divide the
full-sky    into    a    set     of    small    patches    with sizes  $\sim
3.6^{\circ}\times3.6^{\circ}$, and  then randomly select  $35$ patches
in total to  cover a field of  $\sim 450$ square degrees to  match the sky
coverage of  KiDS-450.  By setting a limiting  magnitude of
$\sim 24.5^{{\rm  th}}$ mag in the $r$-band, the  effective number density
in our light-cone is $n \sim 8\ {\rm arcmin}^{-2}$, similar to
that  in the  KiDS-450 observations.  The DLS  is much  deeper,  with a
magnitude  $\sim  27^{\rm th}$  in  $r$-band,  producing an  effective
number density  $\sim 11\ {\rm arcmin}^{-2}$ of  the source population
in 5  tomographic bins.  Given that our lensing  simulation is
performed to  redshift $z_{\rm max} \sim  2.0$, we discard  the 5th redshift bin, 
using only tomographic bins 1-4  of DLS, which gives 
$n \sim 8\ {\rm arcmin}^{-2}$.   Thus, a set of mock  galaxy catalogue 
is constructed to mimic  the sky coverage  and galaxy number
density for each of KiDS-450 and DLS.  For each mock we produce 100
realizations by  randomly sampling the patches at different positions   
to  estimate  the   uncertainties  of   tomographic  shear
correlations due to the cosmic variance and sampling noise.

\subsection{Results}

We    measure   the
auto-correlation and cross-correlation functions $\xi^{(ij)}_{\pm}$ using
the                  public                 code                  {\tt
  Athena}   \footnote{\url{http://www.cosmostat.org/software/athena}},
which  estimates the  second-order  shear  correlation functions  from
equation~(\ref{equ:xipm}).  The  superscript $(ij)$  denotes different
redshift  bins  used for  the  calculation of  the
correlation function. In our case there are four redshift bins labeled
from $1$ to $4$ with increasing redshift.

\subsubsection{Model predictions and comparison with observations}
\label{sec:modelcomparison}

\begin{figure*}[t]
  \centering
  \includegraphics[width=1.98\columnwidth]{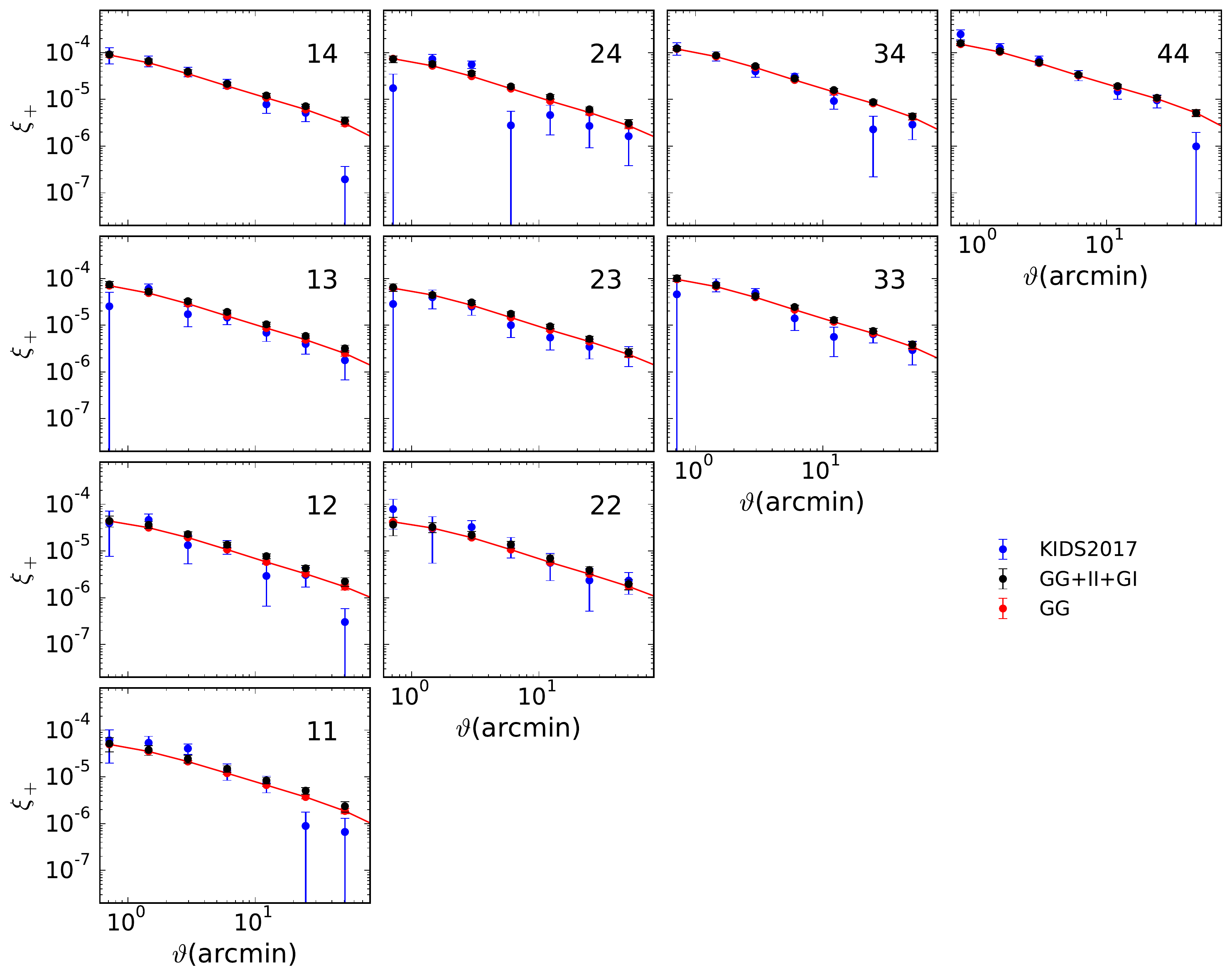}
  \caption{Tomographic measurements of $\xi_{+}$ from our mock KiDS-450
  catalogue with galaxy number density $n = 8.0\ {\rm arcmin}^{-2}$ and
  total sky coverage of $\sim 450$ square degrees. Red circles represent the shear
  correlations from the gravitational field (GG), and black circles are the
  total shear correlations (GG+II+GI), which can be directly compared with
  the observational results (blue circles) from the KiDS-450 \citep{2017MNRAS.465.1454H}.}
  \label{fig7}
\end{figure*}

\begin{figure*}[t]
  \centering
  \includegraphics[width=1.98\columnwidth]{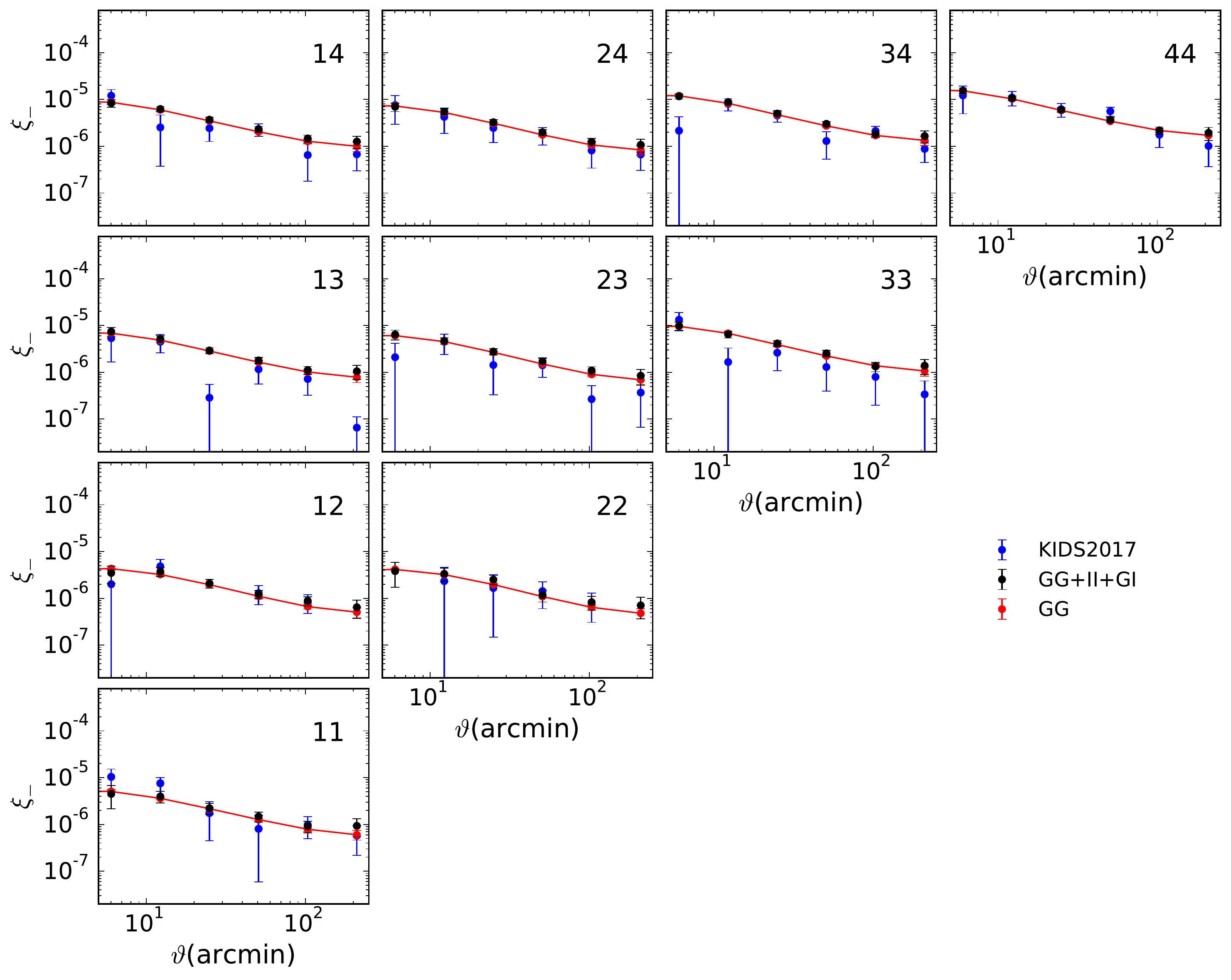}
  \caption{Similar to figure~\ref{fig7}, but for the tomographic
  measurements of $\xi_{-}$.}
  \label{fig8}
\end{figure*}

\begin{figure*}[t]
  \centering
  \includegraphics[width=1.98\columnwidth]{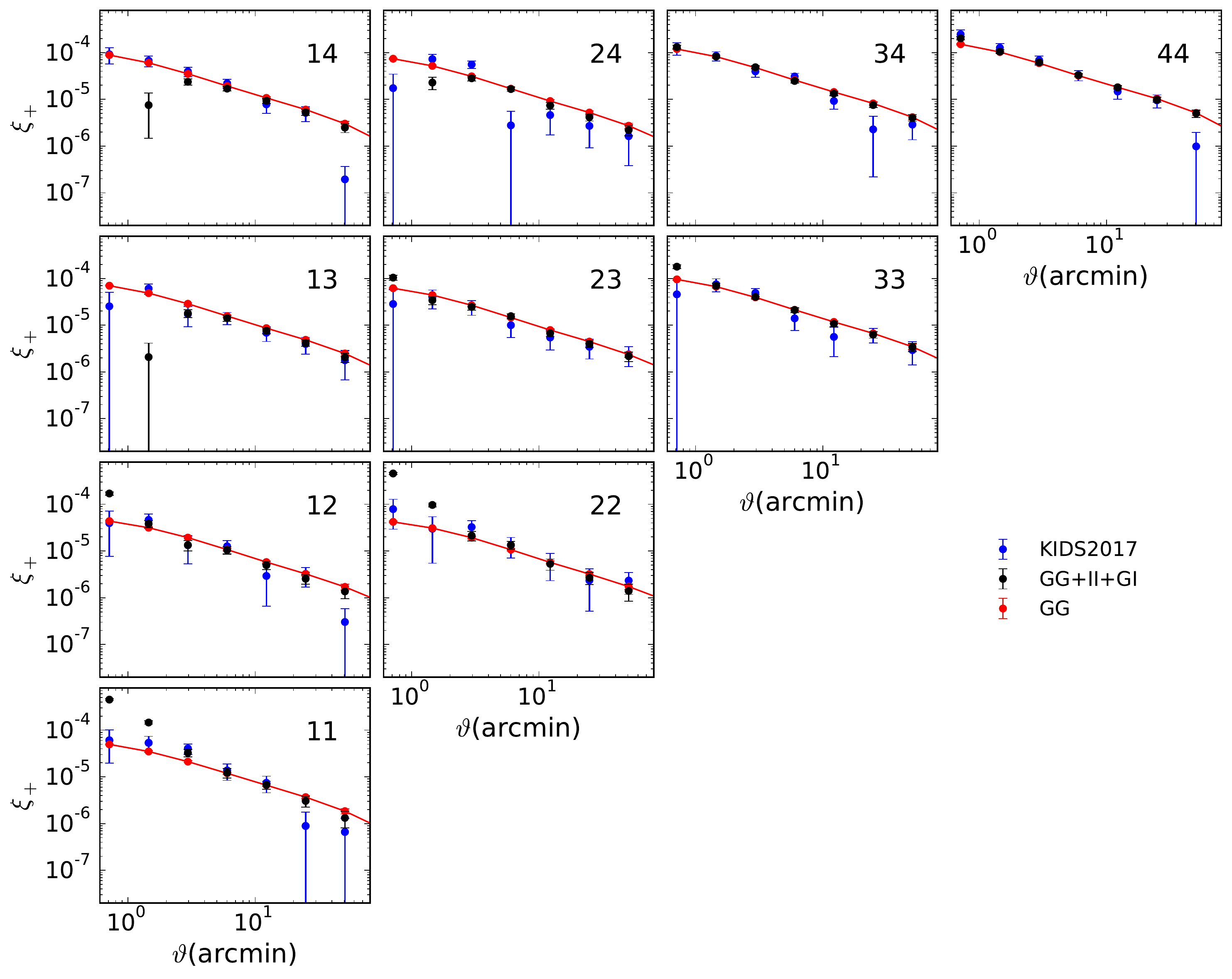}
  \caption{Similar to Fig.~\ref{fig7}, but here 
  satellite galaxies are assumed to be radially aligned
  with the central galaxy while in Fig.~\ref{fig7} satellites are
  assumed to have random orientations. See the text for
  details of how we assign galaxy shape and orientation.}
  \label{fig9}
\end{figure*}

\begin{figure*}
  \centering
  \includegraphics[width=1.98\columnwidth]{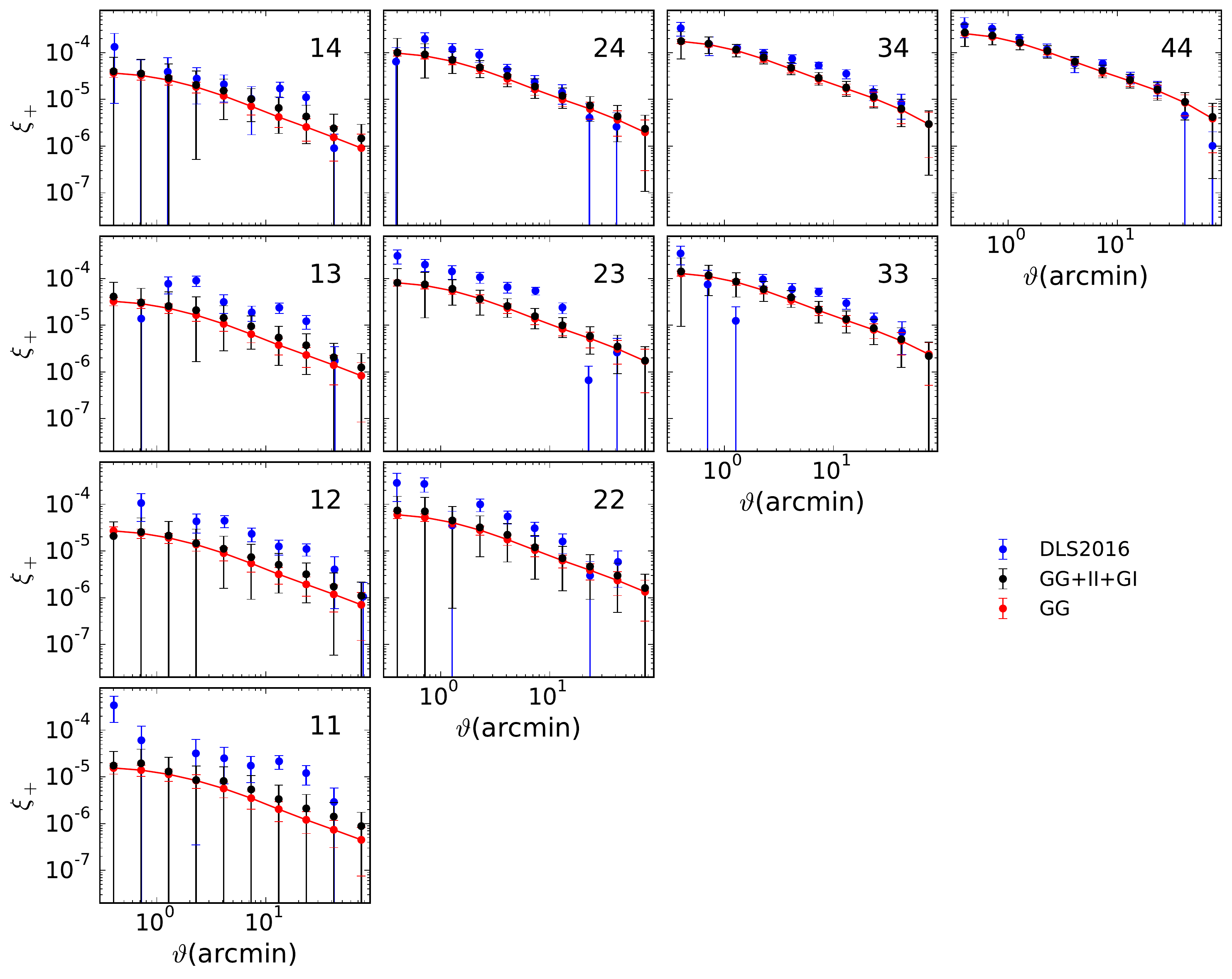}
  \caption{Tomographic shear correlations for the DLS mock catalogue with galaxy
  density $n=8.0\ {\rm arcmin}^{-2}$ and total sky coverage $\sim 20$ square degrees.
  It is seen that the data points are higher than our model predictions.}
  \label{fig10}
\end{figure*}

\begin{figure*}
  \centering
  \includegraphics[width=1.98\columnwidth]{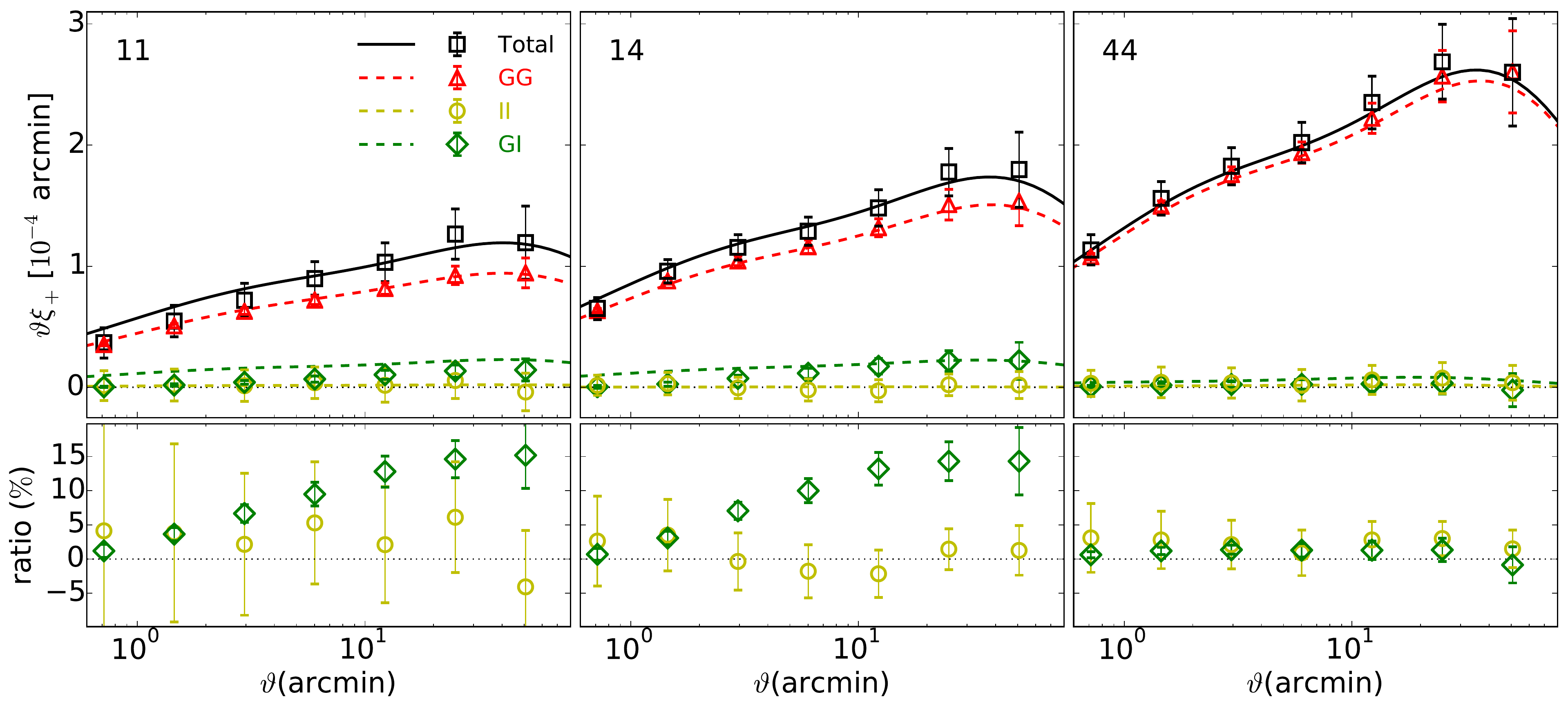}
  \caption{Top panels show the contributions of
  different components to the shear correlations in 
  a few redshift bins, as obtained
  from our mocked KiDS-450. The lower
  panels show the ratios of II and GI to the real
  gravitational shear correlation (GG). In the
  top panels, solid lines show the results obtained from
  the best fit to the non-linear alignment model (Eq.~\ref{equ:IA}).
  Dashed lines show the contributions of GG, II and GI
  in the model fitting.}
  \label{fig11}
\end{figure*}

\begin{figure*}
  \centering
  \includegraphics[width=1.98\columnwidth]{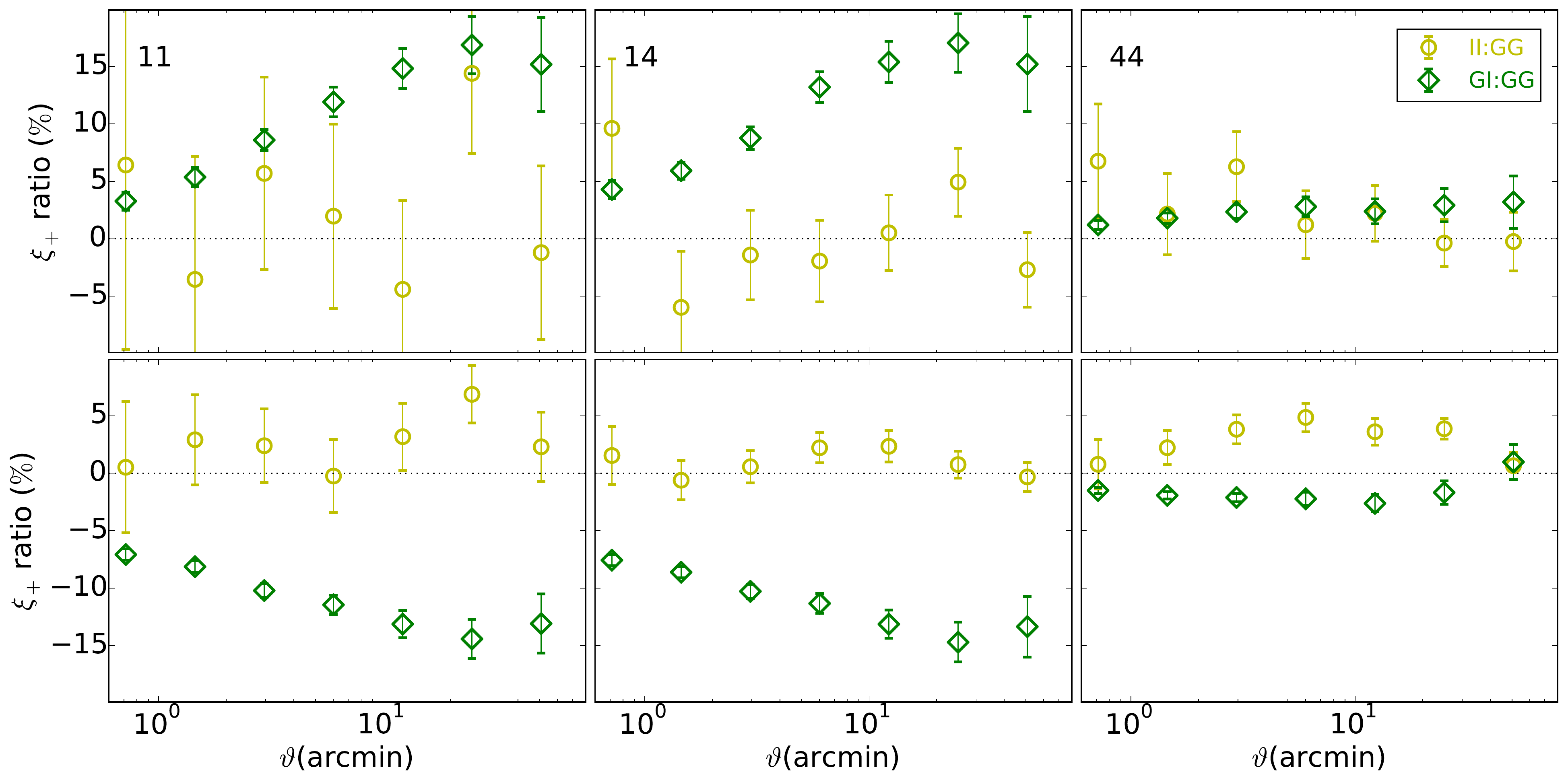}
  \caption{The ratios of the II and GI terms with the gravitational
   shear correlation (GG) in two ideal cases.
  Top  panel:  all  central  galaxies  are  assumed
  to be spirals, and their spins are assumed to follow that of the
  dark matter haloes. Bottom panel: all central galaxies
  are assumed to be ellipticals, and their shapes
  are assumed to follow their host haloes but with a 
  mis-alignment that is given by a
  Gaussian distribution with a dispersion of $35^\circ$.
  It is clearly seen that the GI contribution is positive in the
  first case, but negative in the second.}
  \label{fig12}
\end{figure*}

In Figs.~\ref{fig7} and \ref{fig8} we show  the tomographic shear
correlations  $\xi_{\pm}$  from our  model  and  compare  them with  the
KiDS-450 results. We note that here the
orientation of  the central galaxy is assumed to follow that of the dark  matter halo.
Namely for an elliptical central, its major axis follows that of the halo,
while for spiral central its orientation is determined by halo spin. For
satellite  galaxies,   regardless  of  ellipticals   or  spirals,  their
orientations  are randomly  distributed on the sky.  The black  circles are  the
model predictions for the ellipticity correlation, which  can be directly compared with
the data  (blue circles). As  mentioned in Section 2.3,  the predicted
ellipticity  correlations are combinations  of the  GG,   GI  and  II
correlations. For  simplicity, here we only  show the GG  terms as the red
circles connected  by the  red lines.  We will show the contributions of 
the II and GI terms in  Section 5.3.2.

Figs.~\ref{fig7} and \ref{fig8} show that  in general
the  model predictions (black  circles) agree  well with  the KiDS-450
results. To quantify the difference between the  model and the
data, we calculate the reduced $\chi^{2}$ defined by
\begin{equation} \label{eq:eq29}
  \chi^{2}_{\pm} = \frac{1}{n}\sum\frac{(\xi^{\rm M}_{\pm}-\xi^{\rm D}_{\pm})^{2}}{(\sigma^{\rm D}_{\pm})^{2}}.
\end{equation}
Here $n$ is the number of data points in the tomographic measurements;
$\xi^{\rm  M}_{\pm}$  and  $\xi^{\rm  D}_{\pm}$  represent  the
predicted and observed  tomographic correlations, respectively; 
$\sigma^{\rm D}_{\pm}$  is the  error in  the data.
The error bars we predicted only contain intrinsic ellipticity 
dispersion of galaxy, cosmic variance and shot noise, and they represent 
the dispersion between the results of our 100 realizations, i.e. the 
uncertainties of one realization. The uncertainties of $\xi_{\pm}^{\rm M}$
which we predicted (the mean of 100 realization) are very low, so we  do
not take them into account in Eq.~\ref{eq:eq29}.
We then find the reduced $\chi^{2}_{+(-)}  = 1.70\ (1.82)$ between  
our model prediction and  the KiDS-450  results.
If the correlation between $\xi_{+}$ and $\xi_{-}$ is taken into account,
we should estimate the reduced $\chi^{2}$ from the correlation matrix,
\begin{equation} \label{eq:eq30}
  \chi^{2} = \frac{1}{n}\sum_{i,j}\Delta \xi_{i}C_{ij}^{-1} \Delta \xi_{j},
\end{equation}
where $C_{ij}$ is the covariance matrix of the data \citep{2017MNRAS.465.1454H};
$\Delta \xi_{i}$ is the difference between the model prediction and the data
in the $i$-th separation bin. This gives a reduced $\chi^{2}=1.36$ for the full
data vector of KiDS-450, which is slightly higher than the reduced $\chi^{2}=1.33$ 
in the fiducial analysis of KiDS-450 \citep{2017MNRAS.465.1454H}. 
If we calculate the reduced $\chi^{2}$ for $\xi_{+}$
and $\xi_{-}$ separately, we find $\chi^{2} = 1.23$ and $1.61$, respectively.

The  good  agreement  between  our model  and the KiDS-450 data is
encouraging,  as  this is the  first  time to reproduce the observational 
results using lensed images of  mock galaxies in  N-body simulations 
combined with a realistic model  of galaxy formation. However,  
inspecting Fig.~\ref{fig7} carefully, one can see that for some bins, 
such as 12, 24 and 34, the model predictions are slightly higher than the data on large
scales. Note that the error bars are also larger in these bins and
the data  points are not  well described by the best  fitting model of
\citet{2017MNRAS.465.1454H}.  To further quantify the systematic deviation between
our model prediction and the data, we define the weighted mean deviation as,
\begin{equation}\label{eq:eq31}
  \Delta_{\rm m} = \frac{1}{n}\sum\frac{\Delta \xi_{i}}{\sigma_{i}}.
\end{equation}
Here only errors from the data are used  as the weight $\sigma_{i}$,
{and the correlation between the measurements is not accounted for.} 
{Note that the correlation between the measurements will sightly reduce 
the deviation, similar to previous analyses for the reduced $\chi^{2}$.} 
The systematic bias between the simulations and the observed data
can then be estimated in terms of the standard deviation as
\begin{equation}\label{eq:eq32}
  S = \frac{\Delta_{\rm m}}{\sigma_{\rm m}},
\end{equation}
where $\sigma_{\rm m} = 1/\sqrt{n}$ is the scatter of $\Delta_{\rm m}$.
In this way, we find that the  systematic bias between our model
prediction and KiDS-450 result is $S = 1.80$ and $1.92$ for $\xi_{+}$
and $\xi_{-}$, respectively.
This positive deviation suggests that our predicted correlations
are  slightly,  but  systematically  higher  than  the  KiDS-450 data with a significance of $\sim 1.8 \sigma$.
As the error  bars are often  correlated between  different  redshift
bins,  our analyses of these derivations  might be too simplistic. 
However, we do  not intend to quantify the difference between the  model
and the data in detail, but would like to point out that such  a difference
could be due to the cosmological parameter, 
$\sigma_{8}\sqrt{\Omega_{\rm m}/0.3}=0.79$, adopted in our simulation,
which is slightly larger than that derived one, $0.745 \pm 0.039$, from the KiDS-450 data
\citep{2017MNRAS.465.1454H}.

The results shown in Figs.~\ref{fig7}  and \ref{fig8}  assume
that    satellites     have    random     orientations.     In
\citet{2013MNRAS.436..819J}  the  orientations  of
satellite  galaxies are assumed to be radially  aligned with  
central galaxies. Thus, for an early-type satellite, its major axis is 
assumed to point towards to the central galaxy, while for a late-type satellite, 
its spin is assumed to be perpendicular to the line connecting the satellite to 
the central galaxy. In Fig.~\ref{fig9}  we compare the model results obtained 
by assuming such radial alignments to the  KiDS-450 results.  
We can see that the shear correlations in diagonal panels at lower redshift 
are much higher  than the data on  small scales, except for the highest redshift, 
4-4 bin, where the model prediction  is close to the data.  This indicates
that  the  radial  alignment  model  of satellites  produces  too strong
correlation on small  scales. In addition, this model  also leads to a
strong  negative GI  term, suppressing  the measured  total  signal on
small  scales in  the  cross-correlation  of shears,  as  seen in  the
off-diagonal  bins  (13,  14,  24).
This strong positive correlation at small scales in the
auto-correlation bins, and the strong negative correlation in 
the cross bins can be explained by the balance between the contributions 
of the II term and the GI term. With the J13 assumption for satellites, 
the II term is positive and GI term is negative. In the auto-correlation bins, 
the II term is stronger than the GI term, so as to make a strong positive correlation 
of total signal at small scales, as shown in Fig.~\ref{fig9}. On the other hand, 
in the cross-correlation bins, the contribution of the II term is reduced
relative to the GI term, so that  the correlation at the small scales shows
a strong negative correlation in the cross-correlation bins. 
The reduced $\chi^{2}$ can be calculated from the covariance matrix of the 
data, and we find $\chi^{2} = 3.73$ for the full data vector of KiDS.
Calculating the $\chi^{2}$ separately for $\xi_{+}$ and $\xi_{-}$,
we get $\chi^{2} = 4.63$ and $2.76$, respectively.
The  results  in  Fig.~\ref{fig9}, therefore,  suggest that the radial alignment  
model for satellite galaxies can be rejected. In what follows,   we will only show 
model predictions in which satellites are assumed to have random orientations.


DLS  is  another  weak  lensing survey completed  recently
\citep{2013ApJ...765...74J,                   2016ApJ...824...77J,
  2016AAS...22730707J}.   Compared  to  KiDS-450,  DLS  has  a  smaller  sky
coverage of 20  square degrees. We produce mock  DLS catalogues following
its sky coverage, galaxy  number density and redshift distribution
of  source galaxies  (see Fig.~\ref{fig:KIDSandDLSphotoz}).
Our  model   predictions  and  comparisons   with  DLS  are   shown  in
Fig.~\ref{fig10}.  
Considering that the data of $\xi_{-}$ is not available for DLS \citep{2016ApJ...824...77J},
here we  only  present the  results of  tomographic
correlations  $\xi_{+}$. It  is  seen that  the  model results  (black
circles) are lower than the DLS data (blue circles), especially in the
lower redshift bins. Note  that the error bars in the model
is  slightly  larger.   For  a DLS-like  survey,   the reduced 
$\chi_{+}^{2} = 1.56$,  and   the systematic bias between
the simulation and the data is   $S = -2.57$ for $\xi_{+}$.
The lower reduced $\chi_{+}^{2}$ seems  to indicate  that the
agreement between our  model and DLS is  slightly better than the
agreement with  KiDS-450.  However, it is  clear 
that the  lower reduced $\chi_{+}^{2}$ is also related
to the fact  the  error-bars in DLS data are much larger 
than those of the KiDS. The large error bars in the DLS data
are partly due to its small sky coverage and the smaller sample 
of galaxies. In the tomographic analysis the KiDS-450 sample
is more than 10 times as large as DLS in terms of the total number 
of galaxies. The strong negative systematic bias $S=-2.57$ indicate that
the DLS data are systematically higher than our model predictions.
Since the cosmic variance has been taken into account in the error bar of DLS, 
such a large systematic deviation can hardly be explained by the cosmic variance only.

It is unclear  what causes the discrepancy ($\sim 2\sigma$ in $S_{8}$) between the
observational results of DLS and KiDS-450.  One potential cause 
might be from the estimations  of photo-$z$. For  KiDS 
\citep{2017MNRAS.465.1454H},  the DIR method is used to estimate the 
redshift distribution of galaxies, while for DLS  \citet{2016ApJ...824...77J}  
the BPZ method is adopted.   As  briefly  discussed in  \citet{2017MNRAS.465.1454H},  
the $\chi^{2}$ can increase by $\sim 10$ when switching from the DIR 
redshift distribution                 to                 the                BPZ
distribution.  They also argued  that the deeper DLS data is  harder to be calibrated.
It is beyond our scope to discuss the discrepancy between the KiDS-450 and DLS results in detail.
We refer the reader to the paper cited above for more discussions.

\subsubsection{The contributions of II and GI terms}
\label{sec:GIIIterm}

Our previous model  results  in Fig.~\ref{fig7}  show that 
there is a difference around 10 per cent between the GG term and the total shear correlation. 
The  difference is due to a combination  of II and GI  terms.  
In Fig.~\ref{fig11} we show the contributions from the two components 
separately. As some data points are negative, we plot the $\theta\xi_{+}$ in
linear  scales, and  for  clarity,  we do  not  show  the  observational
data. The  top panel  shows our fiducial  results for the KiDS-450  mock in
some  tomographic  redshift  bins  (the same as in
Fig.~\ref{fig7}), and the lower panel  shows the ratios of GI  and II
with the GG term. As one can see from the plot, the II term is very weak, 
consistent with zero. This  is expected because most  galaxies in our  model are late-types
and      their     intrinsic      alignment      is     very      weak
(Fig.~\ref{fig:eta0}). Moreover, the GI term is basically  positive and 
its contribution could  be as large as 15\%  on large  scales. Following  
the procedures  usually  adopted in observational work to determine the free 
parameter $A_{\rm IA}$ \citep[e.g.,][]{2013MNRAS.432.2433H},  we  fit 
the total signals (black circles) from simulation using the non-linear intrinsic 
alignment model  (Eq.~\ref{equ:IA}).
Note that in our calculation, the GG signal is given using the non-linear 
theoretical power spectrum with given cosmological parameters, and the red dashed 
line in Fig.~\ref{fig11} shows that the theoretical prediction agrees well with 
the measured GG term from our simulation. The best fit to total signal (GG+II+GI) 
gives $A_{\rm  IA} =  -0.972 \pm 0.217$.
The  fit to  each component is also shown as the dashed line in the Fig.~\ref{fig11}.

{Compared with the result by \citet{2017MNRAS.465.1454H}, they constrained the 
amplitude of the intrinsic alignment to a positive $A_{\rm IA}=1.10 \pm 0.64$ in their 
fiducial analysis of KiDS-450, which gives a negative GI term in the measurements.}
While a  negative  $A_{\rm IA}$ here indicates  the  contribution  from  GI term  is
positive  and acts to  increase the  overall correlation  signals.  
In fact, a positive GI signal is not surprising and has been reported in 
major weak lensing surveys.  For example, \citet{2008A&A...479....9F} found that
$A_{\rm IA}=-2.2^{+4.6}_{-3.8}$ from  the third-year CFHTLenS data, and
\citet{2013MNRAS.432.2433H} reported that $A_{\rm IA}=-1.18^{+0.96}_{-1.17}$
from  the final  CFHTLenS  data.  \citet{2017MNRAS.465.2033J} found  that
$A_{\rm     IA} = -3.6 \pm 1.6$     from     the   re-analysis of CFHTLenS data.
\citet{2017MNRAS.465.1454H} found a positive GI  term with $A_{\rm IA}=-1.10_{-0.70}^{+0.96}$
for the KiDS-450 data if they used the BPZ method to estimate  galaxy photometric redshift. 
Recently, \citet{2017arXiv170801538T} found, from the DES data,  that  for spiral
galaxies, the  GI term  is also positive  with $A_{\rm IA}=-0.8$  at a
84 per cent confidence level.

Since most galaxies in our  model are late-types, we conclude that the
positive  GI   signal  is  contributed  by  spirals.    As  a  further
test on the contributions from spiral and elliptical galaxies,  
we show, in Fig.~\ref{fig12}, the  ratios of  the
II and GI terms with  the GG term for two ideal cases.   In the top panel,
we assume  that all central galaxies  in our model  are spirals and
their spins  follow the  spins of  their host dark  matter halos.
In  the  bottom  panel,  we assume  all  central  galaxies  are
ellipticals  and their  shapes follow  the  shapes of  the host  halos
defined with the inertia tensor, but with a misalignment given 
by a gaussian distribution with a dispersion of  $35^{\circ}$.  In the two cases all
satellite galaxies are assumed to have random orientations.  
Fig.~\ref{fig12} shows that the GI term is indeed positive on  all scales 
for spiral  galaxies, although their {II contribution} is close to zero.  
For elliptical galaxies, their  II term  is positive and  the GI  term 
is negative.
{Note here the error bars are different for the two different ratios. 
This can be simply explained from Eq.~\ref{eq:2pcf} by considering the noise in the correlation of GI and II terms. 
As show by the GG correlation, the gravitational shear can be accurately measured in the mock. 
So if we consider the noise (${\bm N}$) in the measurement of intrinsic shapes of galaxies, 
GI and II terms can be expressed as 
$\left< {\bm \epsilon}_{i}^{\rm (I)} {\bm g}_{j}\right> + \left< {\bm N}_{i} {\bm g}_{j}\right>$
and 
$\left< {\bm \epsilon}_{i}^{\rm (I)} {\bm \epsilon}_{j}^{\rm (I)}\right> + 2 \left< {\bm \epsilon}_{i}^{\rm (I)} {\bm N}_{j}\right> + \left< {\bm N}_{i} {\bm N}_{j}\right>$, respectively.
The effect of shape noise can contribute the additional correlations as
$\left< {\bm N}_{i} {\bm g}_{j}\right>$ for GI term 
and $2 \left< {\bm \epsilon}_{i}^{\rm (I)} {\bm N}_{j}\right> + \left< {\bm N}_{i} {\bm N}_{j}\right>$ for II term.
Combining with the definition of correlation $\xi_+$, this difference 
can be used to explain the different error bars for the two different ratios.}

A positive GI term from spiral galaxies is not expected from the tidal
field        model.         From  linear theory
\citep[e.g.,][]{2004PhRvD..70f3526H}, the GI  term is found to be  
negative, which has been used  as a  fiducial model  in weak  lensing  data analyses
\citep[e.g.,][]{2011A&A...527A..26J,     2013MNRAS.432.2433H}.     One
important       assumption        in       the       linear       model
is  that the shape  of foreground galaxies are radially aligned  with the 
nearby over-dense region. This is on average true  for elliptical galaxies. 
But for  spiral galaxies, where alignments are mainly determined  by 
angular momenta of  the dark matter halos through 
large-scale tidal field \citep[see][and references therein]{2009IJMPD..18..173S}, 
this may not be true.  Observationally, it is
found  the spins  of  spiral galaxies  tend  to align  with the  nearby
filaments,  but the  short  axes of  ellipticals  are perpendicular  to
filaments \citep{2010MNRAS.408..897J, 2013ApJ...775L..42T}.  Both
hydro-dynamical  simulations  \citep[e.g.,][]{2015MNRAS.448.3391C} and  N-body
simulations \citep[e.g.,][]{2015ApJ...813....6K, 2017MNRAS.468L.123W} also confirmed
such    a    dependence    on    galaxy    types.    In    particular,
\citet{2015MNRAS.454.2736C}    have    found   from    hydro-dynamical
simulations that  spiral galaxies  have a significant  tendency to be
tangentially  aligned with  over-density  regions.  Their figure 10
demonstrates clearly the  alignment  of  spiral   galaxies  around 
over-density regions and the origin of a positive GI term.

Finally we note that  the II and GI terms are close  related to how we
model galaxy shapes  and orientations. In this paper  we simply assume that
spiral galaxies follow the spins  of dark matter halos. However, as
shown in, e.g.,  \citet[][]{2010MNRAS.404.1137B}, galaxy
spins have  a broad distribution  of mis-alignment with dark matter
haloes. This  mis-alignment   will   reduce   the  positive   GI
terms. Furthermore,  the total GI and  II terms in real data 
depend also on the  fraction  of spiral  and  elliptical  galaxies,  
as well  as  on galaxy luminosities and redshift. 
More comprehensive analyses on these factors are needed 
to quantify their impacts on  GI and II terms. 
This paper, which makes use of  both $N$-body simulation
and galaxies from  a semi-analytical model, is  a step towards  this  goal. 
But  here we  only  focus  on a first comparison of the predicted 
shear correlations with the data.
We will present a more comprehensive investigation on the contribution of GI and
II terms in a future paper.

\section{CONCLUSIONS}

It  is well  known  that the  intrinsic  alignment of  galaxy and  its
associated correlation with the gravitational shear is one of the
dominant contaminations to the  weak lensing survey. Numerous  efforts have
been  devoted   to modeling galaxy-galaxy intrinsic  alignment,
gravitational-galaxy intrinsic alignment  and their  impacts on  the measure
cosmic  shear correlation  \citep[for a  review, see][]{2015PhR...558....1T,
2015SSRv..193....1J, 2015SSRv..193..139K, 2015SSRv..193...67K}. One
useful and direct way  to judge  these alignment models  and their impacts  on the
measured galaxy shear correlations is to produce  mocked galaxy images
using  ray-tracing simulations which can be directly compared with the observational data.

In  this work  we make  a first  attempt to  use a  large cosmological
$N$-body simulation,  ELUCID, and  a semi-analytical model  for galaxy
formation, to perform a full-sky  ray tracing, so as to produce mocked
galaxies images and associated  gravitational shear field.  We compare
our results  on the tomographic  shear correlation with data  from two
recent weak-lensing  surveys, KiDS-450 and  DLS. The main  results are
summarized in the following.

To produce galaxies  images on a curved sky, which  is needed for a survey
with  a  large  sky  coverage,  we  follow the  methods  of  \citet{2013MNRAS.435..115B},
and perform a high-spatial and mass resolution ray tracing
with an  iterative scheme of  spherical harmonic analysis.  We compare
the  measured  power  spectrum  of  convergence  and  shear  with  the
analytical Halofit model and the Born approximation.  It is found that
the measured  power spectrums of convergence and  shear E-/B-mode have
good  agreement   with  the  revised   non-linear  Halofit  prediction
\citep{2012ApJ...761..152T}.  The prediction from Born approximation gives
higher  power at  small scales  than ray-tracing  simulation with $> 10\%$ for $\ell  \ge 6000$.

We  follow \citet{2013MNRAS.431..477J, 2013MNRAS.436..819J}  to assign
shapes to  model galaxies.  For  an early-type central galaxy,  its major
axis  is assumed to align  with that  of  the host  dark matter  halo, and  for
late-type central galaxy its spin  follows that of the halo with major
axis  determined by  projecting the  circular  disk on  the sky.   For
early-type  satellite galaxies,  they  are radially  aligned with  the
central galaxy,  and for late-type  satellites their spin lies  in the
plane perpendicular to the radial  direction to the central galaxy.  We
also consider an additional model in which satellite galaxies have
random orientations.  Using this modes for galaxy shapes, we find that
early-type  central  galaxies  have  strong  {intrinsic ellipticity correlation} but
late-type  galaxies  have  very weak  alignment, in  broad  agreement  with
observations.

To compare with the observational data of KiDS-450 and DLS, we produce
mock surveys  by mimicking their  sky coverage, galaxy  number density
and redshift  distribution of  source galaxies. It  is found  that our
model  with random  orientation  of satellites  agrees  well with 
KiDS-450. Using the covariance matrix of the data \citep{2017MNRAS.465.1454H}, 
we can follow the Eq.~\ref{eq:eq30} to give the reduced $\chi^{2} = 1.36$ 
for the full data vector of KiD-450. This reduced $\chi^{2}$ is slightly 
higher than that in the fiducial analysis of KiDS-450, 
where they obtain a reduced $\chi^{2} = 1.33$ \citep{2017MNRAS.465.1454H}.
In addition, we also calculate the reduced $\chi^{2}$ for $\xi_{+}$ and $\xi_{-}$
separately, and we find the reduced $\chi^{2}$ is 1.23 and 1.61 respectively.
To further quantity the difference between our model prediction and the data,
we estimate this systematic bias $S$ by Eq.~\ref{eq:eq32}. The result shows 
that the systematic bias between our model prediction and KiDS-450 result is
$S=1.80$ and $1.92$ for $\xi_{+}$ and $\xi_{-}$, respectively. In other
words, the $\xi_{\pm}$ we predicted are systematically higher than what KiDS measured 
with a significance of $\sim 1.8 \sigma$.

On other hand, considering that the data of $\xi_{-}$ is not available for the DLS data,
we only compare our model prediction with the data in $\xi_{+}$. Following Eq.~\ref{eq:eq29},
we calculate the reduced $\chi_{+}^{2}$ between the simulation and the data is $1.56$ for 
the DLS data, which seems to be acceptable in our work. While we also find a strong 
negative systematic bias $S=-2.57$ between the model prediction and the DLS result,
which indicates that the DLS data are systematically higher than our model predictions.
Since the cosmic variance has been taken into account in the error bar of DLS, such
a large systematic deviation can hardly be explained by the cosmic variance only.
Moreover, assuming  that there is no scatter in the alignment angle,
we rule out the model in which satellite galaxies are radially aligned
with central galaxies, as it produces too strong power on small scales.

We also study the contributions of the  II and GI terms on the total shear correlations.
It is found that the II term is consistent with zero,
as in  our model  most galaxies  are spirals and  they have  very weak
intrinsic  alignment. Most  importantly, we  detect a  positive  GI term
which  is mainly contributed  by spiral  galaxies. The  GI term  can be
up to 15 per cent on large  scales, and so  its effect  on the  total shear
correlation cannot  be neglected. A positive GI term  is a result of
the correlation between the spins  of spirals and the large-scale
structure, where  it is found  that spiral galaxies  are significantly
tangential aligned with the  nearby over-dense regions. 
This alignment is different from that of elliptical galaxies, which are radially
aligned  with  the  over-dense  regions,  and produces  a  negative  GI
term.

Finally, we note that in our simulation the shape orientation of model
galaxies  is determined by  the host  dark matter  halo, which  is 
probably too simplistic. In fact, there should be mis-alignments for both elliptical
and spiral galaxies. Quantifying these mis-alignments and their dependence on
galaxy properties with observations or hydro-dynamical simulations is crucial.
Our results  suggest that an accurate model of  GI term is very important  for weak-lensing survey,
and it must include the dependence on galaxy type.


\acknowledgments
The authors would like to thank M. R. Becker for making \texttt{CALCLENS} available
 and H. Hildebrandt for providing the redshift distribution of KiDS-450. Also thanks
 are given to Catherine Heymans, John Peacock, Yanchuan Cai , Jun Zhang, Zuhui Fan, 
Liping Fu and Xiangkun Liu for useful comments and suggestions. 
We also thank the anonymous referee for constructive reports 
which significantly improve the quality of the paper.
The work is supported by the NSFC (No.11333008), the 973 program (No. 2015CB857003,
2015CB857002, 2013CB834900), the NSFC (11273179, 
11673065, 11273061, 11233005, 11621303, 11522324, 11421303, and 11473053), 
and the NSF of Jiangsu province (No. BK20140050).

\appendix

\section{Spherical lensing simulation}\label{appendix:rt}
Here we briefly summarize our multi-sphere ray-tracing algorithm
on curved sky. We refer the reader to \citet{2008ApJ...682....1D},
 \citet{2009A&A...497..335T}, \citet{2013MNRAS.435..115B} for a
detailed description of the full-sky lensing simulation.
As described by \citet{2013MNRAS.435..115B}, lensing properties
of our mock galaxies can be extracted from those simulations
through an HEALPix grid search method.

\subsection{Full-sky Lensing Potential and Ray-tracing}

After decomposing the light-cone into a set of shells
with the width of $\sim 50h^{-1}{\rm Mpc}$, we can
obtain the surface matter overdensity $\sigma^{(n)}$
of the $n$-th shell by
\begin{equation}
  \sigma^{(n)}(\bm{\theta}^{(n)}) = \int^{\chi_{n+1/2}}_{\chi_{n-1/2}} {\rm d}\chi' \delta(r(\chi')\bm{\theta}^{(n)}, \chi'),
\end{equation}
The convergence field is then given by
\begin{equation}
  \kappa^{(n)}(\bm{\theta}^{(n)})  = W^{(n)}\sigma^{(n)}(\bm{\theta}^{(n)}),
\end{equation}
where the lensing kernel $W^{(n)}$ is defined as
\begin{equation}
  W^{(n)} = \frac{3}{2} \left( \frac{H_{0}}{c} \right)^{2} \Omega_{{\rm m}} \frac{r(\chi_{n})}{a(\chi_{n})}
\end{equation}
By solving Poisson equation, one can obtain the lensing potential in harmonic space
\begin{equation}
  \phi^{(n)}_{\ell m} = -\frac{2}{\ell(\ell+1)} \kappa^{(n)}_{\ell m}.
\end{equation}

In the context of gravitational lensing, the deflection field $\alpha^{(n)}_{\ell m}$
can be derived from the gravitational lensing potential through
equation~(\ref{equ:alpha1}) and light rays can be propagated to
the next shells following \citep{2009A&A...497..335T},
\begin{equation}
  \bm{x}^{(n+1)} = \mathcal{R}(\bm{n}^{(n)}\times\bm{\alpha}^{(n)}, \lVert\bm{\alpha}^{(n)}\rVert)  \bm{x}^{(n)},
\end{equation}
where the rays are initialized at the center of each HEALPix-cell.
The rotation matrix $\mathcal{R}$ defines the propagation direction
between different shells. The lensing distortion matrix $\mathcal{A}^{(n)}$
can be evaluated by \citep{2013MNRAS.435..115B},
\begin{equation}
\begin{aligned}
  \mathcal{A}^{(n+1)}_{ij} & = \left(1-\frac{D^{n}_{0}}{D^{n+1}_{0}}\frac{D^{n+1}_{n-1}}{D^{n}_{n-1}} \right) \mathcal{A}^{(n-1)}_{ij} + \frac{D^{n}_{0}}{D^{n+1}_{0}}\frac{D^{n+1}_{n-1}}{D^{n}_{n-1}} \mathcal{A}^{(n)}_{ij} - \frac{D^{n+1}_{n}}{D^{n+1}_{0}} \mathcal{U}^{(n)}_{ik}\mathcal{A}^{(n)}_{ij},
\end{aligned}
\end{equation}
where the angular diameter distance
 $D^{n+1}_{n} \equiv r(\chi_{n+1}-\chi_{n})$;
 $\mathcal{U}^{(n)}_{ij}$ is the tidal matrix of the $n$-th shell,
 which can be related to the $2$nd derivatives of the lensing
potential, $\mathcal{U}^{(n)}_{ij} = \phi^{(n)}_{,ij}$.

\begin{figure}
  \centering
  \includegraphics[width=0.75\columnwidth]{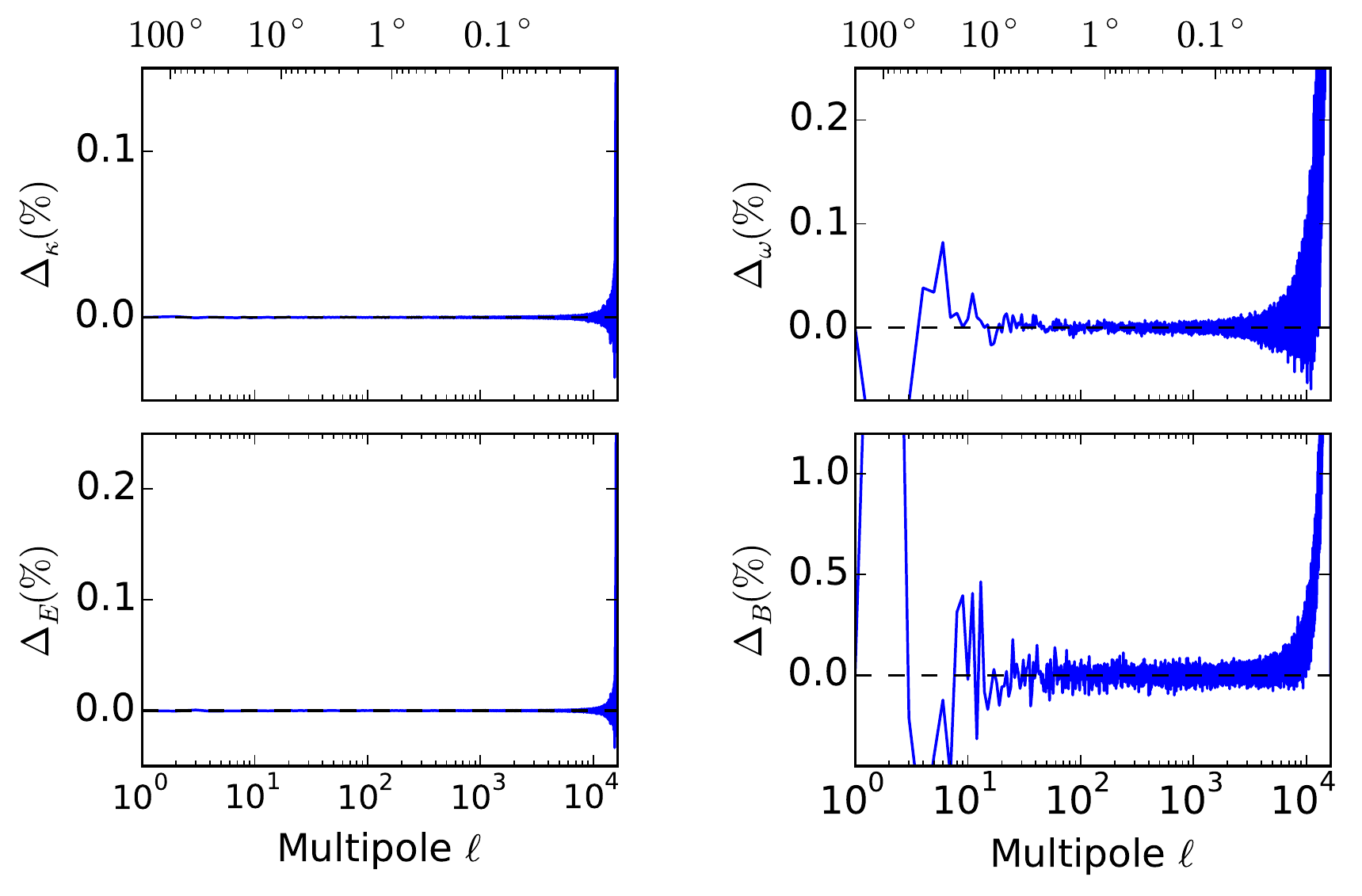}
  \caption{The differences of lensing power spectra measured
  from our ray-tracing simulation with or without the iterative
  algorithm. Here we defined the relative deviation as
  $\Delta = [C_{0}(\ell) - C_{3}(\ell)]/C_{3}(\ell)$, where the
  subscript denotes the order of iteration.}
  \label{fig:dcl2}
\end{figure}

For more accurate computations, we perform
the spherical harmonic analysis with an iterative algorithm,
as is performed in HEALPix subroutine {\tt{map2alm\_iterative}},
to control the residual in the solution of
the lensing potential. The order of iteration in the analysis is 
chosen to be 3, as a compromise with the computational time.
In Figure~\ref{fig:dcl2}, we show the
 statistical measurements with or without this iterative method.
Here we define the relative deviation as
  $\Delta = [C_{0}(\ell) - C_{3}(\ell)]/C_{3}(\ell)$, where the
  subscript denotes the order of iteration.
Clearly, via the iterative algorithm, the numerical errors are
controlled effectively, and the power spectra of shear B-mode is
significantly suppressed, especially for scales smaller
than $\sim1\ {\rm arcmin}$. It is also more accurate than
the multigrid method used in the \texttt{CALCLENS}
\footnote{\url{github.com/beckermr/calclens}} as discussed in \citet{2013MNRAS.435..115B}.

%


\section{Projection of Early-type Galaxy on the Sky}\label{appendix:proj}

As described in \autoref{sec:GM}, the intrinsic shape of early-type galaxies
 can be modeled by the inertial tensor, which defines a triaxial ellipsoid
in the intrinsic reference system as,
\begin{equation}
  x^2 + y^2/p^2 + z^2/q^2 = 1,
\end{equation}
where the axis ratio satisfies $0 < q \leqslant p \leqslant 1$.
We follow \citet{1983Ap&SS..92..335G} to implement the projection onto
the   sky.    The line of sight is defined by the view angle
$(\theta, \varphi)$ in the local coordinate. The axis ratio of the projected ellipse
 on the sky plane can be written as \citep{1977ApJ...213..368S}
\begin{equation}
  r(p,q,\theta, \varphi) = \left[ \frac{A+C-\sqrt{(A-C)^{2} + B^{2}}} {A+C+\sqrt{(A-C)^{2} + B^{2}}} \right]^{1/2},
\end{equation}
where
\begin{equation}
\begin{cases}
A =  q^{2} \sin^2 \theta + (p^{2} \sin^{2} \varphi + \cos^{2} \varphi) \ \cos^{2} \theta\\
B =  (1-p^2) \ \sin 2\varphi \ \cos \theta\\
C =  \sin^{2} \varphi + p^2 \cos^{2} \varphi
\end{cases}
\end{equation}
Finally we change the reference system to the observer's frame  to obtain
the projected ellipses on the full-sky.





\begin{thebibliography}{}
\bibitem[Agustsson \& Brainerd(2010)]{2010ApJ...709.1321A} Agustsson, I., \& Brainerd, T.~G.\ 2010, \apj, 709, 1321
\bibitem[Agustsson \& Brainerd(2006)]{2006ApJ...644L..25A} Agustsson, I., \& Brainerd, T.~G.\ 2006, \apjl, 644, L25
\bibitem[Bailin \& Steinmetz(2005)]{2005ApJ...627..647B} Bailin, J., \& Steinmetz, M.\ 2005, \apj, 627, 647
\bibitem[Baldry et al.(2008)]{2008MNRAS.388..945B} Baldry, I.~K., Glazebrook, K., \& Driver, S.~P.\ 2008, \mnras, 388, 945
\bibitem[Bartelmann \& Schneider(2001)]{2001PhR...340..291B} Bartelmann, M., \& Schneider, P.\ 2001, \physrep, 340, 291
\bibitem[Bartelmann \& Maturi(2016)]{2016arXiv161206535B} Bartelmann, M., \& Maturi, M.\ 2016, arXiv:1612.06535
\bibitem[Battye et al.(2015)]{2015JCAP...04..048B} Battye, R.~A., Moss, A., \& Pearson, J.~A.\ 2015, \jcap, 4, 048
\bibitem[Becker(2013)]{2013MNRAS.435..115B} Becker, M.~R.\ 2013, \mnras, 435, 115
\bibitem[Bett et al.(2007)]{2007MNRAS.376..215B} Bett, P., Eke, V., Frenk, C.~S., et al.\ 2007, \mnras, 376, 215
\bibitem[Bett et al.(2010)]{2010MNRAS.404.1137B} Bett, P., Eke, V., Frenk, C.~S., Jenkins, A., \& Okamoto, T.\ 2010, \mnras, 404, 1137
\bibitem[Binney(1985)]{1985MNRAS.212..767B} Binney, J.\ 1985, \mnras, 212, 767
\bibitem[Blazek et al.(2012)]{2012JCAP...05..041B} Blazek, J., Mandelbaum, R., Seljak, U., \& Nakajima, R.\ 2012, \jcap, 5, 041
\bibitem[Bridle \& King(2007)]{2007NJPh....9..444B} Bridle, S., \& King, L.\ 2007, New Journal of Physics, 9, 444
\bibitem[Bunn(2003)]{2003NewAR..47..987B} Bunn, E.~F.\ 2003, \nar, 47, 987
\bibitem[Bunn et al.(2003)]{2003PhRvD..67b3501B} Bunn, E.~F., Zaldarriaga, M., Tegmark, M., \& de Oliveira-Costa, A.\ 2003, \prd, 67, 023501
\bibitem[Calabretta \& Roukema(2007)]{2007MNRAS.381..865C} Calabretta, M.~R., \& Roukema, B.~F.\ 2007, \mnras, 381, 865
\bibitem[Catelan et al. (2001)]{2001MNRAS.320L...7C}Catelan, P., Kamionkowski, M., \& Blandford R. D.\ 2001, \mnras, 320, 7
\bibitem[Catelan \& Porciani(2001)]{2001MNRAS.323..713C} Catelan, P., \& Porciani, C.\ 2001, \mnras, 323, 713
\bibitem[Chisari et al.(2015)]{2015MNRAS.454.2736C} Chisari, N., Codis, S., Laigle, C., et al.\ 2015, \mnras, 454, 2736
\bibitem[Chisari et al.(2016)]{2016MNRAS.461.2702C} Chisari, N., Laigle, C., Codis, S., et al.\ 2016, \mnras, 461, 2702
\bibitem[Chisari et al.(2017)]{2017MNRAS.472.1163C} Chisari, N.~E., Koukoufilippas, N., Jindal, A., et al.\ 2017, \mnras, 472, 1163
\bibitem[Chisari et al.(2014)]{2014MNRAS.445..726C} Chisari, N.~E., Mandelbaum, R., Strauss, M.~A., Huff, E.~M., \& Bahcall, N.~A.\ 2014, \mnras, 445, 726
\bibitem[Codis et al.(2015)]{2015MNRAS.448.3391C} Codis, S., Gavazzi, R., Dubois, Y., et al.\ 2015, \mnras, 448, 3391
\bibitem[Cole et al.(2000)]{2000MNRAS.319..168C} Cole, S., Lacey, C.~G., Baugh, C.~M., \& Frenk, C.~S.\ 2000, \mnras, 319, 168
\bibitem[Cooray \& Hu(2002)]{2002ApJ...574...19C} Cooray, A., \& Hu, W.\ 2002, \apj, 574, 19
\bibitem[Crittenden et al.(2001)]{2001ApJ...559..552C} Crittenden, R.~G., Natarajan, P., Pen, U.-L., \& Theuns, T.\ 2001, \apj, 559, 552
\bibitem[Dark Energy Survey Collaboration et al.(2016)]{2016MNRAS.460.1270D} Dark Energy Survey Collaboration, Abbott, T., Abdalla, F.~B., et al.\ 2016, \mnras, 460, 1270
\bibitem[Das \& Bode(2008)]{2008ApJ...682....1D} Das, S., \& Bode, P.\ 2008, \apj, 682, 1-13
\bibitem[de Jong et al.(2015)]{2015A&A...582A..62D} de Jong, J.~T.~A., Verdoes Kleijn, G.~A., Boxhoorn, D.~R., et al.\ 2015, \aap, 582, A62
\bibitem[Dong et al.(2014)]{2014ApJ...791L..33D} Dong, X.~C., Lin, W.~P., Kang, X., et al.\ 2014, \apjl, 791, L33
\bibitem[Faltenbacher et al.(2007)]{2007ApJ...662L..71F} Faltenbacher, A., Li, C., Mao, S., et al.\ 2007, \apjl, 662, L71
\bibitem[Faltenbacher et al.(2009)]{2009RAA.....9...41F} Faltenbacher, A., Li, C., White, S.~D.~M., et al.\ 2009, Research in Astronomy and Astrophysics, 9, 41
\bibitem[Foreman et al.(2016)]{2016MNRAS.463.3326F} Foreman, S., Becker, M.~R., \& Wechsler, R.~H.\ 2016, \mnras, 463, 3326
\bibitem[Fosalba et al.(2015)]{2015MNRAS.447.1319F} Fosalba, P., Gazta{\~n}aga, E., Castander, F.~J., \& Crocce, M.\ 2015, \mnras, 447, 1319
\bibitem[Fosalba et al.(2008)]{2008MNRAS.391..435F} Fosalba, P., Gazta{\~n}aga, E., Castander, F.~J., \& Manera, M.\ 2008, \mnras, 391, 435
\bibitem[Fu et al.(2008)]{2008A&A...479....9F} Fu, L., Semboloni, E., Hoekstra, H., et al.\ 2008, \aap, 479, 9
\bibitem[G{\'o}rski et al.(2005)]{2005ApJ...622..759G} G{\'o}rski, K.~M., Hivon, E., Banday, A.~J., et al.\ 2005, \apj, 622, 759
\bibitem[Galletta(1983)]{1983Ap&SS..92..335G} Galletta, G.\ 1983, \apss, 92, 335
\bibitem[Guo et al.(2013)]{2013MNRAS.428.1351G} Guo, Q., White, S., Angulo, R.~E., et al.\ 2013, \mnras, 428, 1351
\bibitem[Hao et al.(2011)]{2011ApJ...740...39H} Hao, J., Kubo, J.~M., Feldmann, R., et al.\ 2011, \apj, 740, 39
\bibitem[Heavens et al.(2000)]{2000MNRAS.319..649H} Heavens, A., Refregier, A., \& Heymans, C.\ 2000, \mnras, 319, 649
\bibitem[Heymans et al.(2013)]{2013MNRAS.432.2433H} Heymans, C., Grocutt, E., Heavens, A., et al.\ 2013, \mnras, 432, 2433
\bibitem[Heymans et al.(2012)]{2012MNRAS.427..146H} Heymans, C., Van Waerbeke, L., Miller, L., et al.\ 2012, \mnras, 427, 146
\bibitem[Heymans et al.(2006)]{2006MNRAS.371..750H} Heymans, C., White, M., Heavens, A., Vale, C., \& van Waerbeke, L.\ 2006, \mnras, 371, 750
\bibitem[Higuchi \& Shirasaki(2016)]{2016MNRAS.459.2762H} Higuchi, Y., \& Shirasaki, M.\ 2016, \mnras, 459, 2762
\bibitem[Hilbert et al.(2009)]{2009A&A...499...31H} Hilbert, S., Hartlap, J., White, S.~D.~M., \& Schneider, P.\ 2009, \aap, 499, 31
\bibitem[Hilbert et al.(2017)]{2017MNRAS.468..790H} Hilbert, S., Xu, D., Schneider, P., et al.\ 2017, \mnras, 468, 790
\bibitem[Hildebrandt et al.(2017)]{2017MNRAS.465.1454H} Hildebrandt, H., Viola, M., Heymans, C., et al.\ 2017, \mnras, 465, 1454
\bibitem[Hinshaw et al.(2013)]{2013ApJS..208...19H} Hinshaw, G., Larson, D., Komatsu, E., et al.\ 2013, \apjs, 208, 19
\bibitem[Hirata \& Seljak(2004)]{2004PhRvD..70f3526H} Hirata, C.~M., \& Seljak, U.\ 2004, \prd, 70, 063526
\bibitem[Hu(2000)]{2000PhRvD..62d3007H} Hu, W.\ 2000, \prd, 62, 043007
\bibitem[Huang et al.(2016)]{2016MNRAS.463..222H} Huang, H.-J., Mandelbaum, R., Freeman, P.~E., et al.\ 2016, \mnras, 463, 222
\bibitem[Jain et al.(2000)]{2000ApJ...530..547J} Jain, B., Seljak, U., \& White, S.\ 2000, \apj, 530, 547
\bibitem[Jee et al.(2016a)]{2016ApJ...824...77J} Jee, M.~J., Tyson, J.~A., Hilbert, S., et al.\ 2016a, \apj, 824, 77
\bibitem[Jee et al.(2013)]{2013ApJ...765...74J} Jee, M.~J., Tyson, J.~A., Schneider, M.~D., et al.\ 2013, \apj, 765, 74
\bibitem[Jee et al.(2016b)]{2016AAS...22730707J} Jee, M.~J., Tyson, J.~A., Hilbert, S., et al.\ 2016b, American Astronomical Society Meeting Abstracts, 227, 307.07
\bibitem[Jing(2002)]{2002MNRAS.335L..89J} Jing, Y.~P.\ 2002, \mnras, 335, L89
\bibitem[Joachimi et al.(2011)]{2011A&A...527A..26J} Joachimi, B., Mandelbaum, R., Abdalla, F.~B., \& Bridle, S.~L.\ 2011, \aap, 527, A26
\bibitem[Joachimi et al.(2013a)]{2013MNRAS.431..477J} Joachimi, B., Semboloni, E., Bett, P.~E., et al.\ 2013a, \mnras, 431, 477
\bibitem[Joachimi et al.(2013b)]{2013MNRAS.436..819J} Joachimi, B., Semboloni, E., Hilbert, S., et al.\ 2013b, \mnras, 436, 819
\bibitem[Joachimi et al.(2015)]{2015SSRv..193....1J} Joachimi, B., Cacciato, M., Kitching, T.~D., et al.\ 2015, \ssr, 193, 1
\bibitem[Jones et al.(2010)]{2010MNRAS.408..897J} Jones, B.~J.~T., van de Weygaert, R., \& Arag{\'o}n-Calvo, M.~A.\ 2010, \mnras, 408, 897
\bibitem[Joudaki et al.(2017)]{2017MNRAS.465.2033J} Joudaki, S., Blake, C., Heymans, C., et al.\ 2017, \mnras, 465, 2033
\bibitem[Kang et al.(2007)]{2007MNRAS.378.1531K} Kang, X., van den Bosch, F.~C., Yang, X., et al.\ 2007, \mnras, 378, 1531
\bibitem[Kang \& Wang(2015)]{2015ApJ...813....6K} Kang, X., \& Wang, P.\ 2015, \apj, 813, 6
\bibitem[Keres et al.(2003)]{2003ApJ...582..659K} Keres, D., Yun, M.~S., \& Young, J.~S.\ 2003, \apj, 582, 659
\bibitem[Kiessling et al.(2015)]{2015SSRv..193...67K} Kiessling, A., Cacciato, M., Joachimi, B., et al.\ 2015, \ssr, 193, 67
\bibitem[Kilbinger(2015)]{2015RPPh...78h6901K} Kilbinger, M.\ 2015, Reports on Progress in Physics, 78, 086901
\bibitem[Kilbinger(2003)]{2003astro.ph..9482K} Kilbinger, M.\ 2003, arXiv:astro-ph/0309482
\bibitem[Kilbinger et al.(2017)]{2017arXiv170205301K} Kilbinger, M., Heymans, C., Asgari, M., et al.\ 2017, arXiv:1702.05301
\bibitem[King \& Schneider(2003)]{2003A&A...398...23K} King, L.~J., \& Schneider, P.\ 2003, \aap, 398, 23
\bibitem[Kirk et al.(2010)]{2010MNRAS.408.1502K} Kirk, D., Bridle, S., \& Schneider, M.\ 2010, \mnras, 408, 1502
\bibitem[Kirk et al.(2015)]{2015SSRv..193..139K} Kirk, D., Brown, M.~L., Hoekstra, H., et al.\ 2015, \ssr, 193, 139
\bibitem[Kitching et al.(2016)]{2016arXiv161104954K} Kitching, T.~D., Alsing, J., Heavens, A.~F., et al.\ 2016, arXiv:1611.04954
\bibitem[Komatsu et al.(2011)]{2011ApJS..192...18K} Komatsu, E., Smith, K.~M., Dunkley, J., et al.\ 2011, \apjs, 192, 18
\bibitem[Krause et al.(2016)]{2016MNRAS.456..207K} Krause, E., Eifler, T., \& Blazek, J.\ 2016, \mnras, 456, 207
\bibitem[Laureijs et al.(2011)]{2011arXiv1110.3193L} Laureijs, R., Amiaux, J., Arduini, S., et al.\ 2011, arXiv:1110.3193
\bibitem[Lemos et al.(2017)]{2017JCAP...05..014L} Lemos, P., Challinor, A., \& Efstathiou, G.\ 2017, \jcap, 5, 014
\bibitem[Levy \& Brustein(2009)]{2009JCAP...06..026L} Levy, D., \& Brustein, R.\ 2009, \jcap, 6, 026
\bibitem[Li \& White(2009)]{2009MNRAS.398.2177L} Li, C., \& White, S.~D.~M.\ 2009, \mnras, 398, 2177
\bibitem[Li et al.(2005)]{2005ApJ...635..795L} Li, G.-L., Mao, S., Jing, Y.~P., et al.\ 2005, \apj, 635, 795
\bibitem[Li et al.(2012)]{2012ApJ...761..151L} Li, M., Pan, J., Gao, L., et al.\ 2012, \apj, 761, 151
\bibitem[Li et al.(2016)]{2016RAA....16..130L} Li, S.-J., Zhang, Y.-C., Yang, X.-H., et al.\ 2016, Research in Astronomy and Astrophysics, 16, 130
\bibitem[Li et al.(2013)]{2013ApJ...768...20L} Li, Z., Wang, Y., Yang, X., et al.\ 2013, \apj, 768, 20
\bibitem[Ling et al.(2015)]{2015PhRvD..92f4024L} Ling, C., Wang, Q., Li, R., et al.\ 2015, \prd, 92, 064024
\bibitem[LSST Science Collaboration et al.(2009)]{2009arXiv0912.0201L} LSST Science Collaboration, Abell, P.~A., Allison, J., et al.\ 2009, arXiv:0912.0201
\bibitem[Luo et al.(2016)]{2016MNRAS.458..366L} Luo, Y., Kang, X., Kauffmann, G., \& Fu, J.\ 2016, \mnras, 458, 366
\bibitem[Mandelbaum et al.(2011)]{2011MNRAS.410..844M} Mandelbaum, R., Blake, C., Bridle, S., et al.\ 2011, \mnras, 410, 844
\bibitem[Mandelbaum et al.(2006)]{2006MNRAS.367..611M} Mandelbaum, R., Hirata, C.~M., Ishak, M., Seljak, U., \& Brinkmann, J.\ 2006, \mnras, 367, 611
\bibitem[Mellier(1999)]{1999ARA&A..37..127M} Mellier, Y.\ 1999, \araa, 37, 127
\bibitem[Meylan et al.(2006)]{2006glsw.conf.....M} Meylan, G., Jetzer, P., North, P., et al.\ 2006, Saas-Fee Advanced Course 33: Gravitational Lensing: Strong, Weak and Micro,
\bibitem[Miyazaki et al.(2012)]{2012SPIE.8446E..0ZM} Miyazaki, S., Komiyama, Y., Nakaya, H., et al.\ 2012, \procspie, 8446, 84460Z
\bibitem[Okumura et al.(2009)]{2009ApJ...694..214O} Okumura, T., Jing, Y.~P., \& Li, C.\ 2009, \apj, 694, 214
\bibitem[Parry et al.(2009)]{2009MNRAS.396.1972P} Parry, O.~H., Eke, V.~R., \& Frenk, C.~S.\ 2009, \mnras, 396, 1972
\bibitem[Peacock \& Smith(2014)]{2014ascl.soft02032P} Peacock, J.~A., \& Smith, R.~E.\ 2014, Astrophysics Source Code Library, ascl:1402.032
\bibitem[Pereira \& Kuhn(2005)]{2005ApJ...627L..21P} Pereira, M.~J., \& Kuhn, J.~R.\ 2005, \apjl, 627, L21
\bibitem[Pereira et al.(2008)]{2008ApJ...672..825P} Pereira, M.~J., Bryan, G.~L., \& Gill, S.~P.~D.\ 2008, \apj, 672, 825-833
\bibitem[Sch{\"a}fer(2009)]{2009IJMPD..18..173S} Sch{\"a}fer, B.~M.\ 2009, International Journal of Modern Physics D, 18, 173
\bibitem[Schneider \& Bridle(2010)]{2010MNRAS.402.2127S} Schneider, M.~D., \& Bridle, S.\ 2010, \mnras, 402, 2127
\bibitem[Schneider et al.(2013)]{2013MNRAS.433.2727S} Schneider, M.~D., Cole, S., Frenk, C.~S., et al.\ 2013, \mnras, 433, 2727
\bibitem[Schneider et al.(2002)]{2002A&A...396....1S} Schneider, P., van Waerbeke, L., Kilbinger, M., \& Mellier, Y.\ 2002, \aap, 396, 1
\bibitem[Schneider et al.(2002)]{2002A&A...389..729S} Schneider, P., van Waerbeke, L., \& Mellier, Y.\ 2002, \aap, 389, 729
\bibitem[Schneider et al.(1992)]{1992grle.book.....S} Schneider, P., Ehlers, J., \& Falco, E.~E.\ 1992, Gravitational Lenses, XIV, 560 pp.~112 figs..~Springer-Verlag Berlin Heidelberg New York.~ Also Astronomy and Astrophysics Library, 112
\bibitem[Shirasaki et al.(2015)]{2015MNRAS.453.3043S} Shirasaki, M., Hamana, T., \& Yoshida, N.\ 2015, \mnras, 453, 3043
\bibitem[Sif{\'o}n et al.(2015)]{2015A&A...575A..48S} Sif{\'o}n, C., Hoekstra, H., Cacciato, M., et al.\ 2015, \aap, 575, A48
\bibitem[Singh et al.(2015)]{2015MNRAS.450.2195S} Singh, S., Mandelbaum, R., \& More, S.\ 2015, \mnras, 450, 2195
\bibitem[Siverd et al.(2009)]{2009arXiv0903.2264S} Siverd, R.~J., Ryden, B.~S., \& Gaudi, B.~S.\ 2009, arXiv:0903.2264
\bibitem[Smith et al.(2003)]{2003MNRAS.341.1311S} Smith, R.~E., Peacock, J.~A., Jenkins, A., et al.\ 2003, \mnras, 341, 1311
\bibitem[Springel(2010)]{2010ARA&A..48..391S} Springel, V.\ 2010, \araa, 48, 391
\bibitem[Springel(2005)]{2005MNRAS.364.1105S} Springel, V.\ 2005, \mnras, 364, 1105
\bibitem[Springel et al.(2005)]{2005Natur.435..629S} Springel, V., White, S.~D.~M., Jenkins, A., et al.\ 2005, \nat, 435, 629
\bibitem[Springel et al.(2001)]{2001MNRAS.328..726S} Springel, V., White, S.~D.~M., Tormen, G., \& Kauffmann, G.\ 2001, \mnras, 328, 726
\bibitem[Stark(1977)]{1977ApJ...213..368S} Stark, A.~A.\ 1977, \apj, 213, 368
\bibitem[Stebbins(1996)]{1996astro.ph..9149S} Stebbins, A.\ 1996, arXiv:astro-ph/9609149
\bibitem[Takahashi et al.(2012)]{2012ApJ...761..152T} Takahashi, R., Sato, M., Nishimichi, T., Taruya, A., \& Oguri, M.\ 2012, \apj, 761, 152
\bibitem[Tempel \& Libeskind(2013)]{2013ApJ...775L..42T} Tempel, E., \& Libeskind, N.~I.\ 2013, \apjl, 775, L42
\bibitem[Tenneti et al.(2016)]{2016MNRAS.462.2668T} Tenneti, A., Mandelbaum, R., \& Di Matteo, T.\ 2016, \mnras, 462, 2668
\bibitem[Tenneti et al.(2014)]{2014MNRAS.441..470T} Tenneti, A., Mandelbaum, R., Di Matteo, T., Feng, Y., \& Khandai, N.\ 2014, \mnras, 441, 470
\bibitem[Teyssier et al.(2009)]{2009A&A...497..335T} Teyssier, R., Pires, S., Prunet, S., et al.\ 2009, \aap, 497, 335
\bibitem[Troxel \& Ishak(2015)]{2015PhR...558....1T} Troxel, M.~A., \& Ishak, M.\ 2015, \physrep, 558, 1
\bibitem[Troxel et al.(2017)]{2017arXiv170801538T} Troxel, M.~A., MacCrann, N., Zuntz, J., et al.\ 2017, arXiv:1708.01538
\bibitem[Tweed et al.(2017)]{2017arXiv170403675T} Tweed, D., Yang, X., Wang, H., et al.\ 2017, arXiv:1704.03675
\bibitem[Van Waerbeke et al.(2001)]{2001A&A...374..757V} Van Waerbeke, L., Mellier, Y., Radovich, M., et al.\ 2001, \aap, 374, 757
\bibitem[Velliscig et al.(2015)]{2015MNRAS.454.3328V} Velliscig, M., Cacciato, M., Schaye, J., et al.\ 2015, \mnras, 454, 3328
\bibitem[Wang et al.(2014)]{2014ApJ...794...94W} Wang, H., Mo, H.~J., Yang, X., Jing, Y.~P., \& Lin, W.~P.\ 2014, \apj, 794, 94
\bibitem[Wang et al.(2016)]{2016ApJ...831..164W} Wang, H., Mo, H.~J., Yang, X., et al.\ 2016, \apj, 831, 164
\bibitem[Wang \& Kang(2017)]{2017MNRAS.468L.123W} Wang, P., \& Kang, X.\ 2017, \mnras, 468, L123
\bibitem[White \& Vale(2004)]{2004APh....22...19W} White, M., \& Vale, C.\ 2004, Astroparticle Physics, 22, 19
\bibitem[Yang et al.(2006)]{2006MNRAS.369.1293Y} Yang, X., van den Bosch, F.~C., Mo, H.~J., et al.\ 2006, \mnras, 369, 1293
\bibitem[York et al.(2000)]{2000AJ....120.1579Y} York, D.~G., Adelman, J., Anderson, J.~E., Jr., et al.\ 2000, \aj, 120, 1579
\bibitem[Zhao \& Baskaran(2010)]{2010PhRvD..82b3001Z} Zhao, W., \& Baskaran, D.\ 2010, \prd, 82, 023001
\bibitem[Zwaan et al.(2005)]{2005MNRAS.359L..30Z} Zwaan, M.~A., Meyer, M.~J., Staveley-Smith, L., \& Webster, R.~L.\ 2005, \mnras, 359, L30




\end{thebibliography}
\end{document}